\RequirePackage{ifpdf}
\ifpdf
\else
\PassOptionsToPackage{dvipdfmx}{graphicx}
\PassOptionsToPackage{dvipdfmx}{hyperref}
\fi

\documentclass[12pt,a4paper]{article}
\usepackage{jheppub}
\allowdisplaybreaks[4]
\usepackage{amsmath,bm}
\usepackage{subfigure}
\usepackage{slashed}
\usepackage{mathrsfs}

\usepackage{color}

\newcommand{\up}{|\!\uparrow\,\rangle}
\newcommand{\dw}{|\!\downarrow\,\rangle}
\newcommand{\upbra}{\langle\,\uparrow\!|}

\newcommand{\pmat}[1]{\begin{pmatrix} #1 \end{pmatrix}}

\newcommand{\nord}[1]{:\mathrel{#1}:}

\title{
Spin systems as quantum field theories in inflationary universe: A study with Unruh-DeWitt detectors
}

\author[a,1]{Shunichiro Kinoshita%
\note{Present affiliation: Kanazawa Institute of Technology, 7-1 Ohgigaoka, Nonoichi, Ishikawa 921-8501, Japan}}
\author[a]{Keiju Murata}
\author[a,b]{Daisuke Yamamoto}
\author[c,d]{Ryosuke Yoshii}
\affiliation[a]{Department of Physics, College of Humanities and Sciences, Nihon University, Sakura-josui,
Tokyo 156-8550, Japan}
\affiliation[b]{RIKEN Center for Quantum Computing (RQC), Wako, Saitama 351-0198, Japan}
\affiliation[c]{Center for Liberal Arts and Sciences, Sanyo-Onoda City University, Yamaguchi 756-0884,
Japan}
\affiliation[d]{International Institute for Sustainability with Knotted Chiral Meta Matter (WPI-SKCM2), Hiroshima University, Higashi-Hiroshima, Hiroshima 739-8526, Japan}
\emailAdd{kinoshita@neptune.kanazawa-it.ac.jp}
\emailAdd{murata.keiju@nihon-u.ac.jp}
\emailAdd{yamamoto.daisuke21@nihon-u.ac.jp}
\emailAdd{ryoshii@rs.socu.ac.jp}

\abstract{%
We propose a method to probe the thermal properties of quantum field theory (QFT) in an inflationary universe simulated by spin systems. Our previous work (arXiv:2410.07587) has demonstrated that QFT of Majorana fermions in an arbitrary two-dimensional spacetime can be mapped onto a spin system. In this study, we apply this mapping to investigate the thermal properties of an inflationary universe. An interaction between a quantum field and a detector allows one to extract information about the quantum field 
from the excitation probability of the detector, known as the Unruh-DeWitt detector. In an inflationary universe with Hubble constant $H$, the excitation probability of an Unruh-DeWitt detector follows a thermal distribution with temperature $H/(2\pi)$, indicating that a static observer in the inflationary universe perceives a thermal field. We consider a spin system corresponding to QFT in an inflationary universe and introduce a single spin interacting with this system as an Unruh-DeWitt detector. We demonstrate that the detector response asymptotically approaches the result of QFT with an appropriate power of the number of spin sites.  
Since the dynamics of spin systems can be implemented on programmable quantum simulation platforms, our study offers a concrete route toward experimentally probing the thermal properties of an inflationary universe in controlled quantum settings. This highlights the potential of quantum technologies to emulate and investigate aspects of quantum field theory in curved spacetimes.
}

\begin{document}
\maketitle

\section{Introduction}

In an inflationary universe, which is well described by de Sitter spacetime, a static observer experiences a fundamental limitation in their observable region due to too rapid cosmic expansion. 
This gives rise to a cosmological event horizon. This horizon, analogous to a black hole horizon, defines a boundary beyond which information cannot reach the observer. The entropy and temperature associated with this horizon can be defined through the first law of thermodynamics, and the observer perceives thermal radiation at a temperature $H / (2\pi)$~\cite{Gibbons:1977mu}, where $H$ is the Hubble constant. This phenomenon, known as the Gibbons-Hawking effect, highlights the thermal nature of quantum fields in curved spacetime. However, due to the extremely small magnitude of such quantum effects, direct observation in our universe is practically infeasible. In this study, we propose a method to investigate the thermal properties of an inflationary universe using spin systems, providing a possible avenue for experimentally probing these fundamental aspects of quantum field theory in curved spacetimes through controllable, tabletop-scale quantum systems.\footnote{As related theoretical proposals , expanding edges of quantum Hall systems have been used to simulate scalar field dynamics in $(1+1)$-dimensional 
Friedmann-Lema\^{\i}tre-Robertson-Walker and de Sitter universes, enabling studies of analog Hawking radiation and entanglement dynamics~\cite{Hotta:2022aiv,Nambu:2023tpg}. }

The interaction between a quantum field and a detector allows one to extract information about the quantum field in curved spacetime from the excitation probability of the detector, known as the Unruh-DeWitt detector~\cite{Unruh1,DeWitt1}. (See also~\cite{BD}.) In an inflationary spacetime with Hubble constant $H$, the excitation probability of an Unruh-DeWitt detector follows a thermal distribution with temperature $H/(2\pi)$, making it possible to characterize the thermal nature of the quantum field through an observable quantity~\cite{Bousso:2001mw}. Our previous work has demonstrated that the theory of Majorana fermions in an arbitrary two-dimensional spacetime can be mapped onto a spin system~\cite{Kinoshita:2024ahu}. In this study, we extend this approach to analyze quantum field theory in an inflationary spacetime using spin systems. In particular, by coupling an Unruh-DeWitt detector to a spin system, we provide a framework in which the excitation probability of a detector in an inflationary universe can be measured experimentally.

The rest of this paper is organized as follows.
In section~\ref{UDdet}, we briefly review the basics of the Unruh-DeWitt detector coupled to quantum fields. 
In particular, we exhibit the relation between the detector response and the Wightman function for quantum fields.
In section~\ref{sec:flat_example}, we consider a quantum field theory in terms of massless Majorana fermion in the two-dimensional flat spacetime and discuss the detector responses.
The main subjects of this paper are discussed in sections~\ref{sec:UDdeSitter} and \ref{sec:spin_deSitter}.
We elucidate the behavior of the Unruh-DeWitt detector in an inflationary universe based on the field theory for the Majorana fermion and the corresponding spin system, respectively.
In section~\ref{single_spin}, we propose a model to implement the Unruh-DeWitt detector in the spin systems.

\section{Unruh-DeWitt detector}
\label{UDdet}

We consider quantum fields in a curved spacetime and a detector interacting with the fields. The Hilbert space of QFT and the detector are denoted as $\mathcal{H}_\textrm{QFT}$ and $\mathcal{H}_\textrm{d}$. The total Hilbert space is given by their tensor product: $\mathcal{H}_\text{tot}=\mathcal{H}_\textrm{QFT} \otimes \mathcal{H}_\textrm{d}$.
The detector is moving in the spacetime with a timelike trajectory $x^\mu=x^\mu(\tau)=(t(\tau),\bm{x}(\tau))$, where $\tau$ is the proper time for the detector. 
In other words, $x^\mu(\tau)$ satisfies $g_{\mu\nu}(x(\tau))\frac{dx^\mu}{d\tau}\frac{dx^\nu}{d\tau}=-1$, where $g_{\mu\nu}(x)$ is the spacetime metric.
In the Schr\"odinger picture, the Hamiltonian describing the time evolution of a coordinate time $t$ is given by
\begin{equation}
    \hat{H}_\textrm{tot}(t)=\hat{H}_\textrm{QFT}(t)\otimes 1 + 1 \otimes \frac{d\tau}{dt} \hat{H}_\textrm{d}+ \lambda \frac{d\tau}{dt} V(t)\ .
    \label{Htot}
\end{equation}
The Hamiltonian of QFT, $\hat{H}_\textrm{QFT}(t)$, can explicitly depend on $t$ through a time dependence of the metric. 
The detector's Hamiltonian, $\hat{H}_\textrm{d}$, is associated with the proper time $\tau$. We assume that it does not depend on time. 
The interaction between the quantum field and the detector, represented by $V(t)$, is written as
\begin{equation}
    V(t)=-h(\tau) \, \Phi(\bm{x}(\tau))\otimes m  \ ,
\end{equation}
where $\Phi$ is a Hermitian operator consisting of the quantum fields in the QFT and $m$ is a Hermitian operator of the detector, independent of time in the Schr\"odinger picture. 
The function $h(\tau)$ is a window function with a compact support in an interval $\tau_0 \leq \tau \leq \tau_1$. 
This describes the switching of the detector, which is switched off outside the interval. 
The $d\tau/dt$ in Eq.~(\ref{Htot}) synchronizes the proper time $\tau$ with the coordinate time of the target spacetime $t$.

We assume that the initial state in $\mathcal{H}_\textrm{QFT}\otimes \mathcal{H}_\textrm{d}$ is 
$|\psi(\tau_0)\rangle = |\Omega\rangle \otimes |E_0\rangle$, where  $|\Omega\rangle$ is the vacuum state in QFT and $|E_0\rangle$ is an energy eigenstate of the detector. Because of the interaction, the detector can be excited (or de-excited) as a result of the time evolution.
The probability that the detector will be excited to a state with an energy $E$ is 
\begin{equation}
    P(E_0\to E) 
    =\lambda^2 |\langle E | m  |E_0\rangle|^2 \mathcal{F}(E-E_0) \ ,
    \label{PE}
\end{equation}
where the detector response function is given by 
\begin{equation}
\begin{split}
    \mathcal{F}(\omega)&=\int^{\tau_1}_{\tau_0}d\tau' \int^{\tau_1}_{\tau_0}d\tau 
    e^{-i\omega (\tau-\tau')} h(\tau) h(\tau') G^+(x(\tau),x(\tau'))\\
    &=\int^{\tau_1}_{\tau_0}d\tau' \int^{\tau_1}_{\tau_0}d\tau 
    e^{i\omega (\tau-\tau')} h(\tau) h(\tau') G^-(x(\tau),x(\tau'))
    \ .
\end{split}
    \label{Fdef}
\end{equation}
See appendix~\ref{derivation_detecorres} for the derivation.
Here, we have introduced the Wightman functions for the vacuum state as
\begin{equation}
    G^+(x,x')=\langle \Omega |\Phi(x)\Phi(x')|\Omega\rangle \ ,\quad 
    G^-(x,x')=\langle \Omega |\Phi(x')\Phi(x)|\Omega\rangle \ ,
\end{equation}
where $\Phi(x)$ is the Heisenberg operator evolved in time by the unperturbed Hamiltonian (i.e., the operator in the interaction picture).
We can say that the Fourier transform of the Wightman function multiplied by window functions represents the detector response.

We can also define the detector response for the thermal state as
\begin{equation}
\begin{split}
    \mathcal{F}_\beta(\omega)&=\int^{\tau_1}_{\tau_0}d\tau' \int^{\tau_1}_{\tau_0}d\tau 
    e^{-i\omega (\tau-\tau')} h(\tau) h(\tau') G^+_\beta(x(\tau),x(\tau'))\\
    &=\int^{\tau_1}_{\tau_0}d\tau' \int^{\tau_1}_{\tau_0}d\tau 
    e^{i\omega (\tau-\tau')} h(\tau) h(\tau') G^-_\beta(x(\tau),x(\tau'))
    \ .
    \end{split}
    \label{Fbetadef}
\end{equation}
This describes the transition probability of the detector interacting with thermal QFT.
Here, we have introduced the thermal Wightman functions as 
\begin{equation}
    G^+_\beta(x,x')=\langle \Phi(x)\Phi(x')\rangle_\beta \ ,\quad 
    G^-_\beta(x,x')=\langle \Phi(x')\Phi(x)\rangle_\beta \ ,
    \label{eq:thermal_Wightman}
\end{equation}
where the thermal average for an operator $\mathcal{O}$ at the inverse temperature $\beta$ is defined by $\langle \mathcal{O}\rangle_\beta = \text{Tr}(e^{-\beta \hat{H}_\textrm{QFT}} \mathcal{O})/Z$, where $Z=\text{Tr}(e^{-\beta \hat{H}_\textrm{QFT}})$.

Now, let us consider the case in which the Wightman functions $G^\pm(x(\tau),x(\tau'))$ depend only on the difference of the proper time $\tau-\tau'$. 
This means that the system has a time translation symmetry along the detector's trajectory, $\tau\to\tau+\textrm{const}$.
We apply the Fourier transformation to $G^\pm$ and $h$ as
\begin{equation}
    G^\pm(x(\tau),x(\tau'))= \int^\infty_{-\infty} \frac{d\omega}{2\pi} \tilde{G}^\pm(\omega) e^{-i\omega(\tau-\tau')}\ ,\quad 
    h(\tau)=\int^\infty_{-\infty} \frac{d\omega}{2\pi} \tilde{h}(\omega) e^{-i\omega \tau}\ .
\end{equation}
Due to time translation symmetry, it can be shown that 
\begin{equation}
    \tilde{G}^-(\omega)=\tilde{G}^+(-\omega)\ .
\end{equation}
From the second line of Eq.~(\ref{Fdef}), the detector response is written as 
\begin{equation}
    \mathcal{F}(\omega)=\int^\infty_{-\infty} \frac{d\omega'}{2\pi} |\tilde{h}(\omega-\omega')|^2 \tilde{G}^-(\omega')
    \ .
    \label{FG}
\end{equation}
Thus, the response function is given by the convolution of the power spectrum of the window function $|\tilde{h}|^2$ and $\tilde{G}^-$. 
We introduce 
the parameter $\mathcal{T}$ defined by  
\begin{equation}
    \mathcal{T}\equiv \int^\infty_{-\infty} d\tau h(\tau)^2  = \int^\infty_{-\infty}\frac{d\omega}{2\pi}|\tilde{h}(\omega)|^2\ ,
    \label{efftime}
\end{equation}
which is referred to as ``effective measurement time''. 
We will focus on the detector response per unit effective measurement time $\mathcal{F}/\mathcal{T}$, which gives a finite value even for the infinite-time measurement, $\mathcal{T}\to \infty$.

One of the simplest choice of the window function is the rectangular window: 
\begin{equation}
    h(\tau)=
    \begin{cases}
        1 & (\tau_0<\tau<\tau_1) \\
        0 & (\textrm{otherwize}) 
    \end{cases}\ ,
    \label{recwindow}
\end{equation}
which means that the detector switches on/off instantaneously.
Then, its Fourier transformation is   
\begin{equation}
    \tilde{h}(\omega)=\frac{2}{\omega}e^{i\omega(\tau_1+\tau_0)/2}\sin \frac{\omega(\tau_1-\tau_0)}{2}\ ,\quad \mathcal{T}=\tau_1-\tau_0\ .
\end{equation}
Note that the effective measurement time $\mathcal{T}$ is nothing but the width of the window function in this case.
If the measurement time becomes infinity, $\mathcal{T}\to\infty$, we obtain $|\tilde{h}(\omega)|^2/\mathcal{T} \to 2\pi \delta(\omega)$. It follows $\mathcal{F}(\omega)/\mathcal{T} \to  \tilde{G}^-(\omega)$ ($\mathcal{T}\to\infty$) from Eq.~(\ref{FG}).

Another choice is the Gaussian window: 
\begin{equation}
    h(\tau)=\exp\left[-\frac{\pi\tau^2}{2\mathcal{T}^2}\right]\ ,
    \label{gausswindow}
\end{equation}
where the parameter $\mathcal{T}$ corresponds to the effective measurement time defined in Eq.~(\ref{efftime}). 
Its Fourier transformation is given by
\begin{equation}
    \tilde{h}(\omega)=\sqrt{2}\mathcal{T} \exp\left[-\frac{\mathcal{T}^2\omega^2}{2\pi}\right]\ .
\end{equation}
Again we have $|\tilde{h}(\omega)|^2/\mathcal{T} \to 2\pi \delta(\omega)$ and $\mathcal{F}(\omega)/\mathcal{T} \to \tilde{G}^-(\omega)$ in the limit of $\mathcal{T}\to\infty$. 
(For the numerical calculations in this paper, in the region where $|\tau| > 3\mathcal{T}$, $h(\tau)$ is forcibly set to $0$, that is, we actually set $\tau_1=-\tau_0=3\mathcal{T}$ and use a compact window function. 
Because the contribution of the integral for $|\tau| > 3\mathcal{T}$ can be neglected practically, it will continue to be ignored in the following discussion.) 

The above results indicate that, if one makes the measurement time sufficiently large, the Fourier transform of the Wightman function can be approximately obtained from the detector response even for either window function. However, in the case of the rectangular window, its power spectrum $|\tilde{h}(\omega)|^2$ exhibits a power-law tail as $\omega\to \infty$, causing contributions from a wide frequency range of the Wightman function to appear in the detector response. In contrast, for the Gaussian window, since its Fourier transform is also Gaussian, the power spectrum of the window function decays exponentially as $\omega\to \infty$. Contributions from only a relatively narrow frequency range of the Wightman function appear in the detector response. 
Therefore, in the case of a finite measurement time, the detector response using a Gaussian window is expected to more clearly extract information about the Wightman function if the Wightman function has a continuous spectrum.

\section{Example of detector response for Majorana fermions in 2D flat spacetime}
\label{sec:flat_example}

\subsection{Majorana fermions in flat spacetime}
As the simplest example, we consider the flat spacetime:
\begin{equation}
    ds^2=-dt^2+dx^2\ ,
\end{equation}
where we assume that the spatial direction is compactified as $x\sim x+\ell$. (In this section, we will eventually consider the infinite spatial region taking the limit $\ell \to \infty$, but in later sections, we use the results for finite $\ell$.)
As QFT, we consider the massless Majorana fermions whose Lagrangian density is given by
\begin{equation}
    \mathcal{L}=-i  \bar{\psi} \slashed{\partial} \psi .
\end{equation}
where $\bar{\psi}=\psi^\dagger\gamma^0$ and $\psi=(\psi_1,\psi_2)^T$ are a two-component spinor field of real Grassmann variables, satisfying $\psi_a^\dagger = \psi_a$ $(a=1,2)$. The Feynman slash notation is defined as  $\slashed{\partial}=\gamma^i \partial_i$ ($i=0,1$) and the gamma matrices $\gamma^i$ in the Majorana representation can be written as
\begin{equation}
\begin{split}
    \gamma^0&=i\sigma^y=
    \begin{pmatrix} 
    0 & 1 \\
    -1 & 0
    \end{pmatrix}\ ,\quad 
    \gamma^1=\sigma^z=
    \begin{pmatrix} 
    1 & 0 \\
    0 & -1
    \end{pmatrix}\ .
\end{split}
\end{equation}
They satisfy $\{\gamma^i,\gamma^j\}=2\eta^{ij}=2 \textrm{diag}(-1,1)$. 
We will assume that the Majorana field satisfies the anti-periodic boundary condition: $\psi(t,x+\ell)=-\psi(t,x)$.

Introducing the complex variable $\Psi=\psi_2-i\psi_1$, we can rewrite the Lagrangian density as
\begin{equation}
    \mathcal{L}=\frac{i}{2}(\Psi^\dagger\partial_t \Psi +\Psi\partial_t \Psi^\dagger)
    -\frac{1}{2}(\Psi\partial_x \Psi -\Psi^\dagger \partial_x \Psi^\dagger)\ .
\end{equation}
The Hamiltonian is simply written as
\begin{equation}
    \hat{H}_\textrm{QFT}=
    \frac{1}{2}\int^{\ell/2}_{-\ell/2} dx(\Psi\partial_x \Psi -\Psi^\dagger \partial_x \Psi^\dagger)\ .
    \label{Hmink}
\end{equation}
The Majorana field satisfies the canonical anti-commutation relations as
\begin{equation}
    \{\Psi(t,x),\Psi^\dagger(t,y)\}=\delta(x-y)\ ,\quad 
    \{\Psi(t,x),\Psi(t,y)\}=\{\Psi^\dagger(t,x),\Psi^\dagger(t,y)\}=0\ .
    \label{canrel}
\end{equation}

\subsection{Solution of the Heisenberg equation}
The Heisenberg equation for $\Psi$ is given by
\begin{equation}
    \dot{\Psi}-i\Psi^\dagger{}'=0\ ,\quad \dot{\Psi}^\dagger+i\Psi'=0\ .
\end{equation}
We apply the Fourier transform as
\begin{equation}
    \Psi(\eta,x)=\frac{1}{\sqrt{\ell}}\sum_{k\in K} e^{ikx} \Psi_k(t)\ ,\quad 
    \Psi^\dagger (\eta,x)=\frac{1}{\sqrt{\ell}}\sum_{k\in K} e^{ikx} \Psi^\dagger_{-k}(t)\ ,
\end{equation}
where, from the anti-periodic boundary condition, the domain of the wave number is
\begin{equation}
K=\left\{\frac{2\pi}{\ell}\left(n-\frac{1}{2}\right)\bigg|n\in \bm{Z}\right\}\ .
\label{Kdef_cont}
\end{equation}
Then, the Heisenberg equations are rewritten as 
\begin{equation}
\left[\frac{d}{dt} + k\pmat{0 & 1\\-1 & 0}\right]\pmat{\Psi_k(t)\\\Psi_{-k}^\dagger(t)} = 0\ .
\end{equation}
Decomposing the time-dependence of variables into the Fourier modes as $(\Psi_k, \Psi_{-k}^\dagger) \propto e^{-i\omega t}$, we obtain eigen-frequencies and their eigen-vectors as
\begin{equation}
    \omega =|k|\ ,\quad \frac{1}{\sqrt{2}}\pmat{1 \\ i\, \textrm{sgn}(k)}\ ,\qquad 
    \omega =-|k|\ ,\quad \frac{1}{\sqrt{2}}\pmat{i\, \textrm{sgn}(k) \\ 1}\ .
\end{equation}
Thus, the general solution of the Heisenberg equation in the momentum space is given by
\begin{equation}
    \pmat{\Psi_k(t)\\\Psi_{-k}^\dagger(t)} = \frac{1}{\sqrt{2}}\left[\gamma_k e^{-i|k|t} \pmat{1 \\ i\,\textrm{sgn}(k)} + \gamma_{-k}^\dagger e^{i|k|t} \pmat{i\,\textrm{sgn}(k) \\ 1}\right]\ .
\end{equation}
In the position space, we obtain
\begin{equation}
\begin{split}
    &\Psi(t,x)=\frac{1}{\sqrt{2\ell}}\sum_{k\in K} e^{ikx} (\gamma_k e^{-i|k|t}+i\,\textrm{sgn}(k) \gamma_{-k}^\dagger e^{i|k|t})\ ,\\
    &\Psi^\dagger(t,x)=\frac{1}{\sqrt{2\ell}}\sum_{k\in K} e^{ikx} (i\,\textrm{sgn}(k) \gamma_k e^{-i|k|t}+\gamma_{-k}^\dagger e^{i|k|t})\ .
\end{split}
\label{Minksol}
\end{equation}
Using the canonical anti-commutation relation~(\ref{canrel}), we obtain 
\begin{equation}
    \{\gamma_k,\gamma_{k'}^\dagger\}=\delta_{k,k'}\ ,\quad 
    \{\gamma_k,\gamma_{k'}\}=\{\gamma_k^\dagger,\gamma_{k'}^\dagger\}=0\ .
    \label{gammacomu}
\end{equation}
Substituting Eq.~(\ref{Minksol}) into Eq.~(\ref{Hmink}), we can write the Hamiltonian of the Majorana fermion as 
\begin{equation}
    H=\sum_{k\in K} |k| \gamma_k^\dagger \gamma_k + E_0\ ,
    \label{Hk}
\end{equation}
where $E_0=-\sum_k |k|/2$ is the vacuum energy. 
Although this summation naively diverges, it can be formally rewritten as $-\sum_k |k|/2 = \zeta(-1) \pi/\ell$, where $\zeta(s)$ is Riemann's zeta function. Thus, we obtain a Casimir energy $E_0 = - \pi/(12\ell)$ after appropriate subtraction procedures. 
The vacuum state $|\Omega\rangle$ is defined by 
\begin{equation}
    \gamma_k |\Omega\rangle =0 \ ,\quad (\forall k\in K)\ .
    \label{vacdef}
\end{equation}
Excited states can be constructed by acting the creation operators on $|\Omega\rangle$.

\subsection{Detector response in flat spacetime}

Now, to discuss the detector response, we consider a detector coupled to the Majorana field through a scalar operator~\cite{Hummer:2015xaa}.
We define the Hermitian scalar operator as
\begin{equation}
    \Phi(t,x)\equiv i\nord{\bar{\psi}(t,x)\psi(t,x)}\ ,
    \label{Phidef}
\end{equation}
where the normal ordering for an operator $\mathcal{O}$ is denoted by $\nord{\mathcal{O}}$. By the normal ordering, all creation operators $\gamma_k^\dagger$ are to the left of all annihilation operators $\gamma_k$ in the operator $\mathcal{O}$, where the spin statistics are taken into account when exchanging the operators. For example, $\nord{\gamma_k \gamma_{k'}^\dagger}=-\gamma_{k'}^\dagger \gamma_k$. 
Using the complex variable $\Psi$, the scalar operator is given by $\Phi=\nord{\Psi^\dagger\Psi}$. The above operator is also written as the 
\begin{equation}
    \Phi(t,x)= i\bar{\psi}(t,x)\psi(t,x)-i\langle \Omega | \bar{\psi}(t,x)\psi(t,x) |\Omega\rangle \ .
    \label{Phidef2}
\end{equation}
In this expression, the divergence of the operator is canceled by its vacuum expectation value.

\begin{figure}[t]
  \centering
\subfigure[Zero temperature]
 {\includegraphics[scale=0.5]{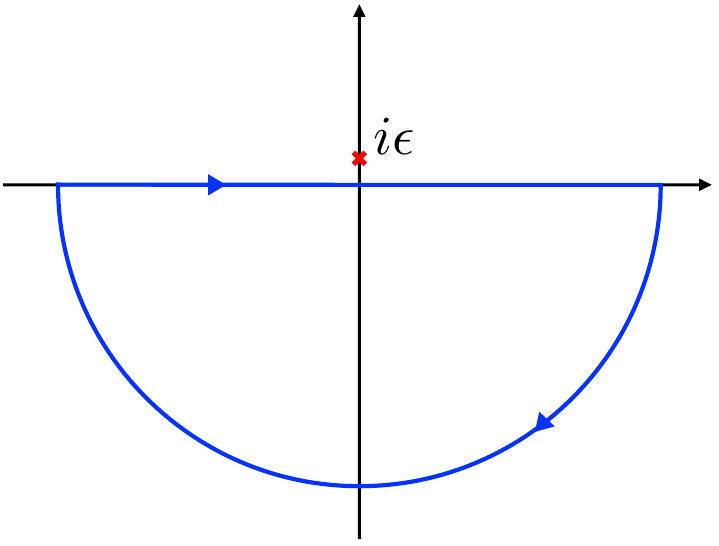}\label{T0countour}
  }
  \subfigure[Finite temperature]
 {\includegraphics[scale=0.5]{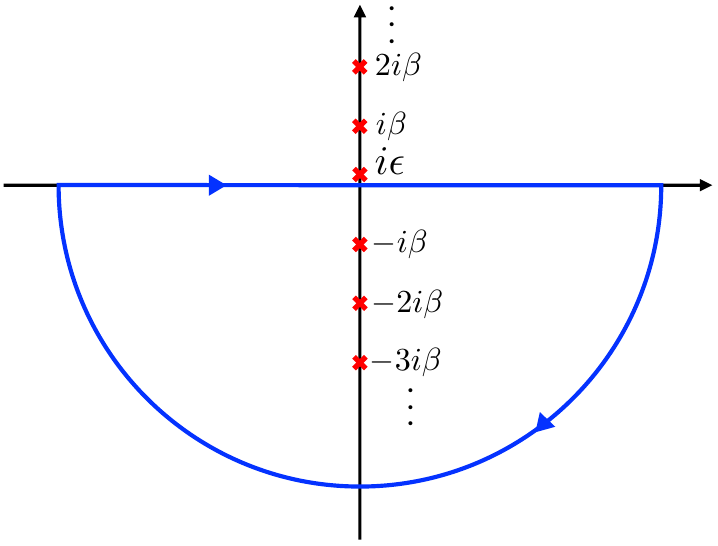}\label{Tfinitecountour}
  }
 \caption{Contours for integrations.
}
 \label{countours}
\end{figure}

As in appendix~\ref{WightmanMink},
for the vacuum state, the Wightman function for $\Phi$ is directly computed as
\begin{equation}
\begin{split}
&G^+(t,x,t',x')
=\langle \Omega|\Phi(t,x)\Phi(t',x')|\Omega\rangle \\
&=-\frac{1}{4\ell^2 \sin[\frac{\pi}{\ell}(t-t'-(x-x')-i\epsilon)]\sin[\frac{\pi}{\ell}(t-t'+x-x'-i\epsilon)]}\ ,
\end{split}
\label{GpMink_ell}
\end{equation}
where $\epsilon$ is a positive infinitesimal.
Especially, in the limit of $\ell\to \infty$, the above expression becomes
\begin{equation}
    G^+(t,x,t',x')=-\frac{1}{4\pi^2((t-t'-i\epsilon)^2-(x-x')^2)}\ .
    \label{GpMink}
\end{equation}
Let us consider a static detector at the origin whose trajectory is given by $(t,x)=(\tau,0)$. 
Then, the Wightman function on the detector's trajectory is 
\begin{equation}
    G^+(t(\tau),0,t(\tau'),0)=-\frac{1}{4\pi^2(\tau-\tau'-i\epsilon)^2}\ .
    \label{Gplus}
\end{equation}
From Eq.~(\ref{Fdef}), for the rectangular window~(\ref{recwindow}) with  $\tau_0\to -\infty$ and $\tau_1\to \infty$, the detector response is 
\begin{equation}
\begin{split}
    \mathcal{F}(\omega)&=-\frac{1}{4\pi^2}\int^{\infty}_{-\infty}d\tau' \int^{\infty}_{-\infty} d\tau
    \frac{e^{-i\omega (\tau-\tau')} }{(\tau-\tau'-i\epsilon)^2}\\
    &=-\frac{1}{4\pi^2}\int^{\infty}_{-\infty}d\tau_c \int^{\infty}_{-\infty} d\Delta \tau
    \frac{e^{-i\omega \Delta \tau}}{(\Delta\tau-i\epsilon)^2}\ ,
\end{split}
\end{equation}
where we have introduced 
\begin{equation}
    \Delta\tau=\tau-\tau'\ ,\quad \tau_c=\frac{\tau+\tau'}{2}\ .
    \label{taucdef}
\end{equation}
The integration over $\tau_c$ gives an infinite factor in the detector response. This is because the Wightman function has time translation symmetry, which results in the number of quanta absorbed by the detector being constant over time. Here, we consider the transition probability per unit time: $\mathcal{F}(\omega)/\mathcal{T}$ with $\mathcal{T}=\int^\infty_{-\infty} d\tau_c$. 
Taking the contour of the integration as in Fig.~\ref{T0countour}, we obtain
\begin{equation}
    \frac{\mathcal{F}(\omega)}{\mathcal{T}}
    =-\frac{1}{4\pi^2}\int^{\infty}_{-\infty} d\Delta \tau
    \frac{e^{-i\omega \Delta \tau}}{(\Delta\tau-i\epsilon)^2} = 0\ .
\end{equation}
This is a trivial result: a static detector in Minkowski spacetime remains unexcited.

Let us next consider the case where the field theory is in a thermal state. the thermal Wightman function for $\ell\to\infty$ is also computed in appendix~\ref{WightmanMink} as
\begin{multline}
     G^+_\beta(t,x,t',x')=\langle \Phi(t,x)\Phi(t',x')\rangle_\beta \\= -\frac{1}{4\beta^2 \sinh\left[\frac{\pi}{\beta} (t-t'-(x-x')-i\epsilon)\right] \sinh\left[\frac{\pi}{\beta} (t-t'+x-x'-i\epsilon)\right]}\ .
\end{multline}
Again we consider the static observer $(t(\tau),x(\tau))=(\tau,0)$. 
Then, the Wightman function on the detector's trajectory is 
\begin{equation}
    G^+_\beta(t(\tau),0,t(\tau'),0)=-\frac{1}{4\beta^2\sinh^2\left[\frac{\pi}{\beta} (\Delta \tau -i\epsilon)\right]}\ .
    \label{Gbeta}
\end{equation}
The detector response is now given by
\begin{equation}
    \frac{\mathcal{F}_\beta(\omega)}{\mathcal{T}}=-\frac{1}{4\beta^2}\int^{\infty}_{-\infty} d\Delta \tau
    \frac{e^{-i\omega \Delta \tau}}{\sinh^2\left[\frac{\pi}{\beta} (\Delta \tau -i\epsilon)\right]}
    =\frac{\omega}{2\pi(e^{\beta\omega}-1)}\ .
    \label{Fbeta}
\end{equation}
This integrand has poles at $\Delta \tau = i\beta n+i\epsilon$ ($n\in\bm{Z})$. 
We have taken the contour of the integration as in Fig.~\ref{Tfinitecountour}.
The transition probability to excited states of the detector is nonzero because of the thermal fluctuation of QFT.
The thermal properties of QFT are reflected in the Boltzmann factor of the detector response.

\subsection{Kubo-Martin-Schwinger relation}

Now, we recall the thermal Wightman functions, which are defined by (\ref{eq:thermal_Wightman}).
Using the cyclic property of the trace, we have 
\begin{equation}
\begin{split}
    G^+_\beta(x,x')&=Z^{-1}\textrm{Tr}(e^{-\beta \hat{H}_\textrm{QFT}} \Phi(x)\Phi(x'))\\
    &=Z^{-1}\textrm{Tr}(e^{-\beta \hat{H}_\textrm{QFT}}\Phi(x')e^{-\beta \hat{H}_\textrm{QFT}} \Phi(x) e^{\beta \hat{H}_\textrm{QFT}}))\\
    &=Z^{-1}\textrm{Tr}(e^{-\beta \hat{H}_\textrm{QFT}}\Phi(x') \Phi(x+i\beta))\\
    &=G^-_\beta(x+i\beta,x')\ ,
    \end{split}
\end{equation}
where $\beta$ in the argument means $\beta \delta^\mu_t$.
This relationship is referred to as the Kubo-Martin-Schwinger (KMS) relation~\cite{Kubo,MS}. This holds for the thermal Wightman function of any operator. Along the trajectory of the static observer, the KMS relation is written as 
$G^+_\beta(\tau, 0, \tau', 0) = G^-_\beta(\tau + i\beta, 0, \tau', 0)$. From its Fourier transform, we obtain
\begin{equation}
    \tilde{G}^-_\beta(\omega)=e^{-\beta \omega} \tilde{G}^+_\beta(\omega)\ .
\end{equation}
As discussed in section~\ref{UDdet}, the detector response will coincide with the Wightman function as $F_\beta(\omega) = \tilde{G}^-_\beta(\omega)=\tilde{G}^+_\beta (-\omega)$ for an infinite measurement time. Thus, the above equation implies
\begin{equation}
    \frac{F_\beta(\omega)}{F_\beta(-\omega)} = e^{-\beta \omega}\ .
    \label{KMSforF}
\end{equation}
It can be immediately verified that Eq.~(\ref{Fbeta}) satisfies this relation.

\subsection{Finite-time measurement}

\begin{figure}[t]
  \centering
\subfigure[Domain of $(\tau_c,\Delta \tau)$]
 {\includegraphics[scale=0.45]{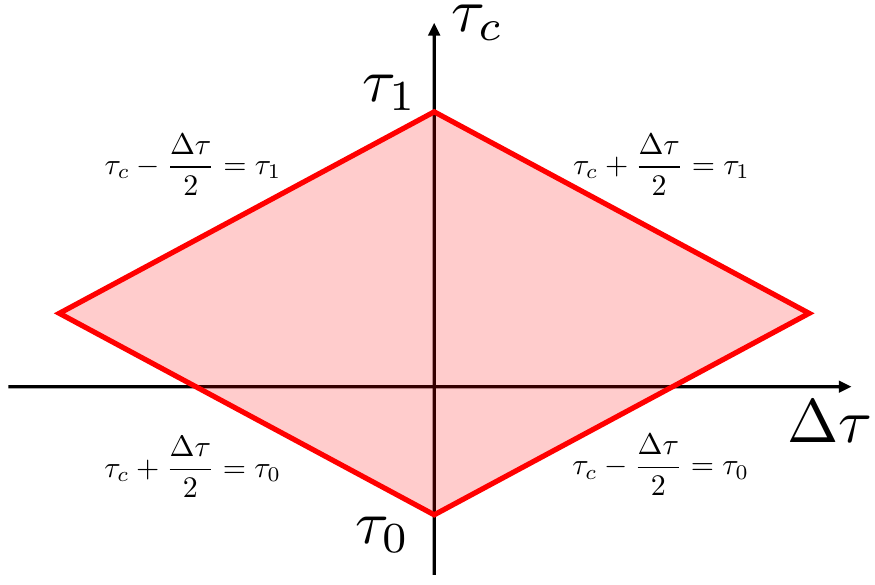}\label{diamond}
  }
  \subfigure[Contour of the numerical integration]
 {\includegraphics[scale=0.55]{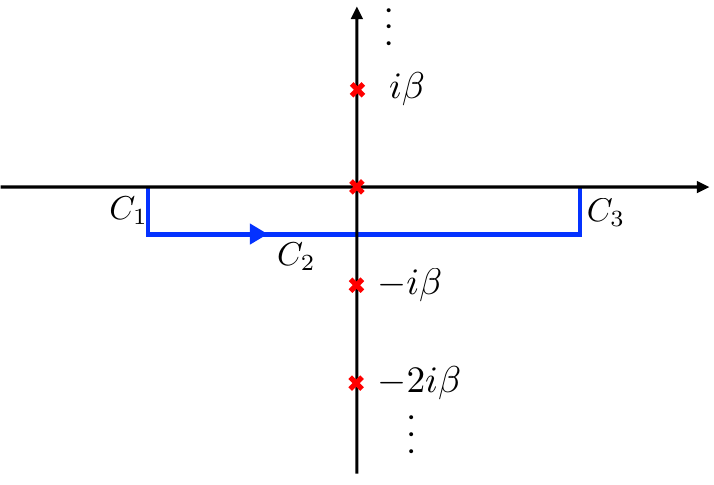}\label{Ftimecontour}
  }
 \caption{
 (a)The domain of $(\tau_c,\Delta \tau)$. Numerical integration is performed over the region shaded in red.
 (b)Contour of the numerical integration for the detector response with a finite-time measurement.
}
\end{figure}

\begin{figure}
\begin{center}
\includegraphics[scale=0.7]{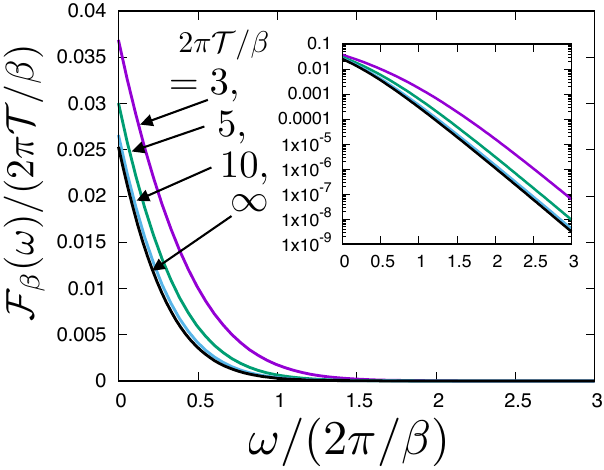}
\end{center}
\caption{Detector response in thermal Minkowski. The time duration of the interaction is varied as $2\pi \mathcal{T}/\beta=3,5,10$ and $\infty$. The inset shows the same figure using a logarithmic scale.
}
\label{Ffintet}
\end{figure}

So far, we have focused on the case of infinite measurement time, where $\tau_0 \to -\infty$ and $\tau_1 \to \infty$ in Eq.~(\ref{Fdef}). However, in the actual experiment, the duration of the measurement should be finite. 
Now, we elucidate the effect of the finiteness of the measurement time on the detector response in the thermal Minkowski. 
As we mentioned previously, switching on the detector coupling for a finite time interval is equivalent to introducing an appropriate window function $h(\tau)$ in the detector response.
Here, we adopt the Gaussian window~(\ref{recwindow}) as the window function.
In terms of $\tau_c$ and $\Delta \tau$ defined in Eq.~(\ref{taucdef}), the domain of integration is the diamond-like region shown in Fig.~\ref{diamond}. It is written as
\begin{equation}
    \tau_0 \leq \tau_c \leq \tau_1\ ,\quad 
    -\tau_\textrm{max} \leq \Delta \tau \leq \tau_\textrm{max}\ .
\end{equation}
where $\tau_\textrm{max}\equiv \textrm{min}(2\tau_c -2\tau_0,-2\tau_c+2\tau_1)$.
The integrand has a pole at $\Delta \tau=0$. So, the original contour of the integration $-\tau_\textrm{max}\to \tau_\textrm{max}$ along the real axis is not suitable for the numerical integration. Thus, we deform the contour into the complex plane as 
\begin{equation}
    C=C_1+C_2+C_3\ .
    \label{contourc123}
\end{equation}
where
\begin{equation}
\begin{split}
    &C_1: -\tau_\textrm{max} \to -\tau_\textrm{max}-iY\ ,\\
    &C_2: -\tau_\textrm{max}-iY \to \tau_\textrm{max}-iY\ ,\\
    &C_3: \tau_\textrm{max}-iY \to \tau_\textrm{max}\ .
\end{split}
\end{equation}
We chose $Y=i\beta/2$ as an offset from the real axis. The contour is shown in Fig.\ref{Ftimecontour}. 
This above contour go through between poles. 
Then, the response function for a finite temperature is written as 
\begin{equation}
    \mathcal{F}_\beta(\omega)=-\frac{1}{4\beta^2}\int_{\tau_0}^{\tau_1} d\tau_c  \int_C dz\,   e^{-i\omega z}h(\tau_c+z/2)h(\tau_c-z/2)\frac{1}{\sinh^2(\pi z/\beta)}\ .
\end{equation}
For the numerical calculation, we simply use the trapezoidal rule.
The result is shown in Fig.~\ref{Ffintet} for $2\pi\mathcal{T}/\beta=3,5,10,\infty$. As the measurement time increases, the detector response approaches the Planckian distribution, which corresponds to an infinite measurement time. The inset displays the same figure using a logarithmic scale. We observe that the detector response has the behavior $\sim e^{-\beta \omega}$ for $\omega \gg 1/\beta$.  This result suggests that the temperature of the system can be measured from the large-$\omega$ behaviour of the detector response even for a finite measurement time.

\section{Unruh-DeWitt detector in the inflationary universe}
\label{sec:UDdeSitter}

\subsection{de Sitter spacetime}

\begin{figure}
\begin{center}
\includegraphics[scale=0.3]{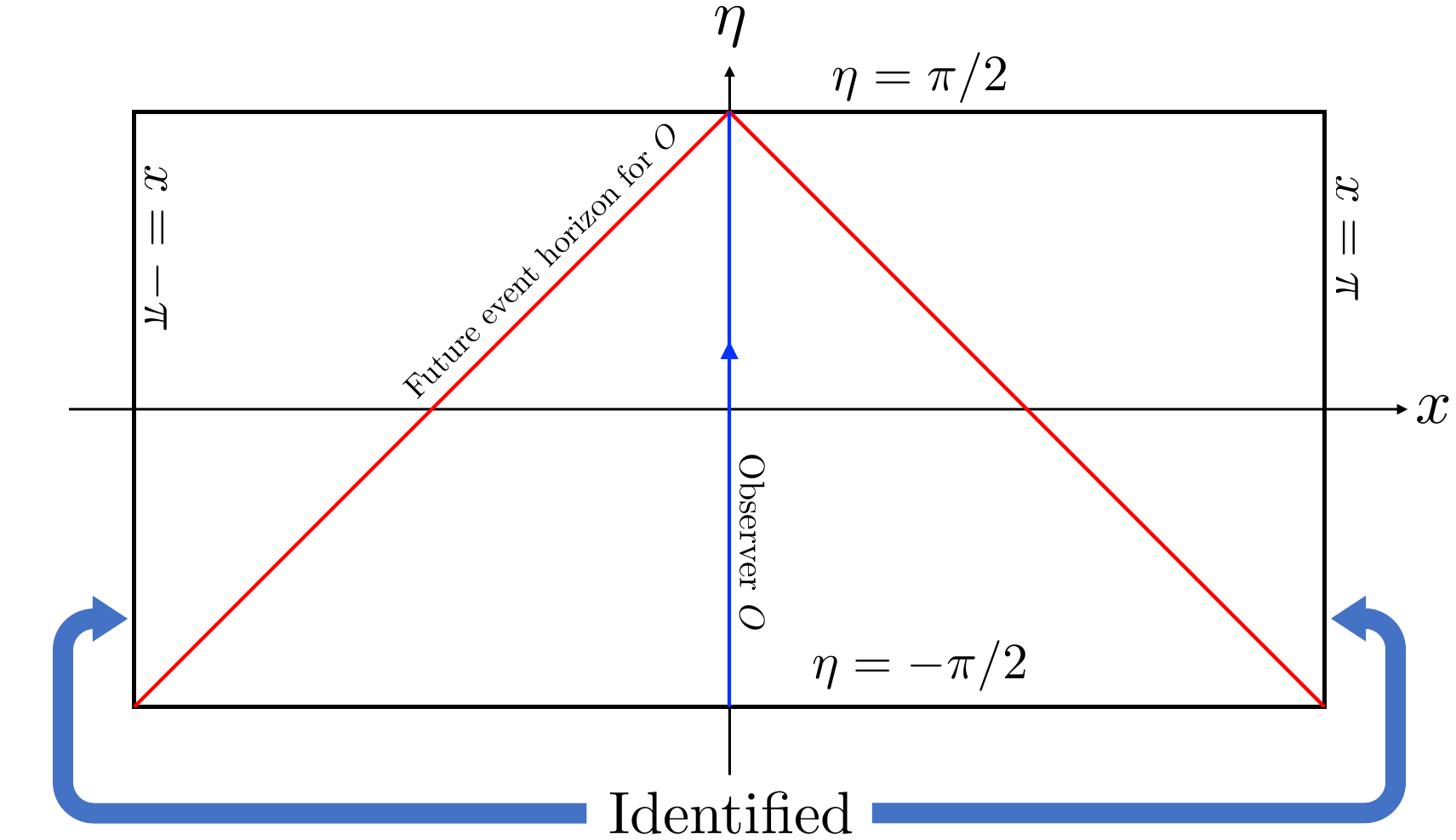}
\end{center}
\caption{
Penrose diagram of the de Sitter spacetime. The red lines represent the future event horizon for a static observer 
$O$ located at the origin $x=0$.
}
\label{dS_diag}
\end{figure}

Let us consider the $(2+1)$-dimensional Minkowski spacetime $\bm{R}^{2,1}$ whose metric is given by $ds^2_3=-(dX^0)^2+(dX^1)^2+(dX^2)^2$. The (1+1)-dimensional de Sitter spacetime is defined as a submanifold of $\bm{R}^{2,1}$:
\begin{equation}
    -(X^0)^2+(X^1)^2+(X^2)^2=\frac{1}{H^2}\ ,
\end{equation}
where $H$ is a constant that corresponds to the Hubble parameter describing the expansion rate. We introduce coordinates of the de Sitter spacetime $(t,x)$ as
\begin{equation}
    X^0=\frac{1}{H}\sinh (H t)\ ,\quad X^1=\frac{1}{H}\cosh (H t) \cos x\ ,\quad X^2=\frac{1}{H}\cosh (H t) \sin x\ ,
\end{equation}
where the spatial coordinate is compactifield as $x \sim x +2\pi$. We will take its domain as $-\pi \leq x \leq \pi$.
Then,  the metric of the de Sitter spacetime is written as
\begin{equation}
    ds^2=-dt^2+\frac{1}{H^2} \cosh^2 (H t)\, dx^2\  \quad (-\infty < t < \infty).
\end{equation}
The coordinates $(t,x)$, which cover the whole de Sitter spacetime with topology $\bm{R}^1\times S^1$, are called global coordinates. 
In order to examine the global causal structure of the de Sitter spacetime, it is useful to introduce the conformal time $\eta$ as $d\eta= H dt/\cosh (Ht)$. The relation between time coordinates $t$ and $\eta$ is explicitly written as 
\begin{equation}
    e^{Ht}=\tan\left[\frac{1}{2}\left(\eta+\frac{\pi}{2}\right)\right]\ .
\end{equation}
In terms of the conformal time $\eta$, the metric of the de Sitter spacetime can be rewritten as the conformally flat form   
\begin{equation}
    ds^2 = \frac{1}{H^2\cos^2 \eta}(-d\eta^2 + dx^2)\ .
\end{equation}
Note that the conformal time is defined in $-\pi/2\leq \eta \leq \pi/2$. The Penrose diagram of the de Sitter spacetime is given in Fig.~\ref{dS_diag}.

Let us focus on an observer $O$ located at the origin $x=0$. Such an observer only perceives the region where $\eta \leq -|x|+\pi/2$. Its boundary $\eta = -|x|+\pi/2$ is called the future event horizon for the observer $O$.

\subsection{Majorana fermions in de Sitter spacetime}

The Lagrangian for the massless Majorana fermion in two-dimensional curved spacetimes is~\cite{Polyakov:1981re,deLacroix:2023uem,Erbin,Kinoshita:2024ahu}
\begin{equation}
    \mathcal{L}=-i \sqrt{-g}\, \bar{\psi} \slashed{\partial} \psi . \label{eq:Majorana_action}
\end{equation}
For the curved spacetime, the Feynman slash notation is defined as $\slashed{\partial} = \gamma^i e_i^\mu  \partial_\mu$ where $e^\mu_i$ is the zweibeins. Latin indices $i, j, k, \ldots$ and Greek indices $\mu, \nu, \rho, \ldots$ are used for the local Lorentz coordinates and the general coordinates, respectively.
The term involving the spin connection vanishes for the ($1+1$)-dimensional Majorana fermions~\cite{deLacroix:2023uem,Erbin,Kinoshita:2024ahu}.
For the de Sitter spacetime, we adopt the following zweibeins:
\begin{equation}
e^0_\mu dx^\mu =\frac{1}{H\cos\eta} d\eta\ ,\quad 
    e^1_\mu dx^\mu =\frac{1}{H\cos\eta}  dx \ .
\label{zwei0}
\end{equation}
The Lagrangian (\ref{eq:Majorana_action}) is rewritten as 
\begin{equation}
    \mathcal{L}=-\frac{i}{H\cos\eta} \bar{\psi} (\gamma^0\partial_\eta + \gamma^1 \partial_x) \psi .
\end{equation}
Introducing the complex variable 
\begin{equation}
    \Psi=\frac{1}{\sqrt{H\cos\eta}}(\psi_2-i\psi_1)\ ,
    \label{Psidef}
\end{equation}
the Hamiltonian 
\begin{equation}
    \hat{H}_\textrm{QFT}=
    \frac{1}{2}\int^{\pi}_{-\pi} dx(\Psi\partial_x \Psi -\Psi^\dagger \partial_x \Psi^\dagger)\ .
    \label{Hds}
\end{equation}
In the conformally flat coordinates $(\eta,x)$, 
it is the same form as the Hamiltonian for the flat spacetime as in Eq.~(\ref{Hmink}).
This is a consequence of the fact that the theory of massless Majorana fermions~(\ref{eq:Majorana_action}) is conformally invariant.
Thus, the solution of the Heisenberg equation is obtained by replacing $t$ with $\eta$ and setting $\ell=2\pi$ in Eq.~(\ref{Minksol}):
\begin{equation}
\begin{split}
    &\Psi(\eta,x)=\frac{1}{\sqrt{4\pi}}\sum_{k\in K} e^{ikx} (\gamma_k e^{-i|k|\eta}+i\,\textrm{sgn}(k) \gamma_{-k}^\dagger e^{i|k|\eta})\ ,\\
    &\Psi^\dagger(\eta,x)=\frac{1}{\sqrt{4\pi}}\sum_{k\in K} e^{ikx} (i\,\textrm{sgn}(k) \gamma_k e^{-i|k|\eta}+\gamma_{-k}^\dagger e^{i|k|\eta})\ ,
\end{split}
\label{dSsol}
\end{equation}
where the domain of the wave number $k$ is given by $K=\{ n-1/2 |n\in \bm{Z}\}$.

\subsection{Detector feels thermal bath in the inflationary universe}

In the de Sitter spacetime, the Bunch-Davis vacuum~\cite{Bunch:1978yq}, $|\Omega\rangle$, is defined by
\begin{equation}
    \gamma_k |\Omega\rangle =0 \ ,\quad (\forall k\in K)\ .
    \label{BDdef}
\end{equation}
In other words, it is a conformal vacuum defined with respect to the conformal Killing vector $\partial_\eta$.
The scalar operator~(\ref{Phidef}), which couples to the detector, is given by
\begin{equation}
    \Phi(\eta,x)=i\nord{\bar{\psi}(\eta,x)\psi(\eta,x)}=H\cos\eta \nord{\Psi^\dagger(\eta,x)\Psi(\eta,x)}\ .
\end{equation}
Since the solution of the Heisenberg equation has the same form as in flat spacetime, the Wightman function for $\nord{\Psi^\dagger(\eta,x)\Psi(\eta,x)}$ is given by Eq.~(\ref{GpMink_ell}) with $\ell=2\pi$ and $t$ replaced by $\eta$.
It follows that the Wightman function for the operator $\Phi$ becomes
\begin{equation}
\begin{split}
    &G^+(\eta,x,\eta',x') =\langle \Omega|\Phi(\eta,x)\Phi(\eta',x')|\Omega\rangle\\
    &=-\frac{H^2 \cos\eta \cos\eta'}{16\pi^2 \sin[\frac{1}{2}(\eta-\eta'-(x-x')-i\epsilon)]\sin[\frac{1}{2}(\eta-\eta'+x-x'-i\epsilon)]}\ .
\end{split}
\end{equation}
We consider that the detector is located at the origin, $x = 0$. Its conformal time $\eta$ is expressed in terms of the proper time $\tau$ through
\begin{equation}
    e^{H\tau}=\tan\left[\frac{1}{2}\left(\eta(\tau)+\frac{\pi}{2}\right)\right]\ .
    \label{taueta}
\end{equation}
Using the above equation, we obtain the Wightman function on the detector's trajectory as 
\begin{equation}
    G^+(\eta(\tau),0,\eta(\tau'),0)=-\frac{H^2}{16\pi^2\sinh^2 \left[\frac{H}{2}(\Delta \tau-i\epsilon)\right]}\ ,
\end{equation}
where $\Delta \tau \equiv \tau - \tau'$ denotes a time interval for the detector's proper time.
This is the same form as Eq.~(\ref{Gbeta}) with $\beta=2\pi/H$. Therefore, we obtain
\begin{equation}
    \frac{\mathcal{F}(\omega)}{\mathcal{T}}
    =\frac{\omega}{2\pi(e^{2\pi \omega/H}-1)}\ .
\end{equation}
This indicates that the detector in the inflationary universe feels the thermal radiation with the temperature $\beta^{-1}=H/(2\pi)$. 

We can also consider the finite-time measurement and obtain completely the same results as in the case of the thermal Minkowski with $\beta^{-1}=H/(2\pi)$.

\section{Spin model for inflation}
\label{sec:spin_deSitter}

\subsection{Spin model for Majorana fermions in de Sitter spacetime}

In Ref.~\cite{Kinoshita:2024ahu}, the spin model Hamiltonians corresponding to the QFT of Majorana fermions in arbitrary two-dimensional curved spacetimes have been revealed.
In particular, if we consider the Friedmann-Lema\^{\i}tre-Robertson-Walker spacetime, in the Hamiltonian with respect to the conformal time $\eta$, the scale factor $a(\eta)$ appears only in the form of a product with the mass $m$ of the Majorana particle. It follows that, for the massless case, regardless of the functional form of the scale factor, the Hamiltonian of the spin system is given as
\begin{equation}
    \hat{H}_\textrm{Spin}=-\frac{1}{2\varepsilon}\sum_{j=1}^L (\sigma^x_j \sigma^x_{j+1} + \sigma^z_j)\ ,
    \label{Hspin}
\end{equation}
where 
\begin{equation}
    \varepsilon=\frac{2\pi}{L}
\end{equation}
represents the lattice spacing. This is nothing but the transverse-field Ising model at the critical point. 
We note that the nature of being a de Sitter spacetime will be taken into account by coupling with a particle detector in section~\ref{subsec:detector_spin}.

The spin operators $\sigma^\pm_j=(\sigma_j^x \pm i \sigma_j^y)$ and $\sigma^z_j$ are written in terms of fermionic operators $c_j$ and $c_j^\dagger$ via the Jordan-Wigner transformation as
\begin{equation}
\begin{split}
    \sigma^+_j=&\prod_{l=1}^{j-1}(1-2c_l^\dagger c_l) c_j\ ,\quad 
    \sigma^-_j=\prod_{l=1}^{j-1}(1-2c_l^\dagger c_l) c_j^\dagger\ ,\\ \sigma_j^z=&1-2c_j^\dagger c_j\ ,
    \label{JW}
\end{split}
\end{equation}
where $c_j$ and $c_j^\dagger$ satisfy the canonical anti-commutation relations $\{c_j,c_l^\dagger\}=\delta_{jl}$ and $\{c_j,c_l\}=\{c_j^\dagger,c_l^\dagger\}=0$. After the Jordan-Wigner transformation, the Hamiltonian~(\ref{Hspin}) becomes 
\begin{equation}
    \hat{H}_\textrm{Spin}=-\frac{1}{2\varepsilon}\sum_{j=1}^L (c_{j+1}c_{j} + c_{j}^\dagger c_{j+1}^\dagger + c_{j}^\dagger c_{j+1}+c_{j+1}^\dagger c_{j} + 1-2c_{j}^\dagger c_{j})\ .
    \label{Hforc}
\end{equation}
Here, we only focus on the state with even fermion number and impose the anti-periodic boundary condition for the fermion operator as $c_{L+1}=-c_1$. 
As shown in Ref.~\cite{Kinoshita:2024ahu}, the above Hamiltonian reduces to that of Majorana fermions~(\ref{Hds}) in the continuum limit $\varepsilon\to 0$, with the following identification:
\begin{equation}
    \Psi(\eta, x_j)= \frac{c_j(\eta)}{\sqrt{\varepsilon}}\ ,\quad x_j=\varepsilon(j-1)\ .
\end{equation}
The coordinates were assigned such that the spatial coordinate of the 
$j=1$ spin is set to $x=0$.

\subsection{Exact solution of Heisenberg equations}

The Heisenberg equations for $c_j$ and $c_j^\dagger$ are exactly solved by
\begin{equation}
\begin{split}
    c_j(\eta)&= \frac{1}{\sqrt{L}}\sum_{\kappa\in \mathcal{K}} e^{i\kappa j}(u_\kappa  \gamma_\kappa  e^{-i\epsilon_\kappa  \eta} + v_\kappa  \gamma_{-\kappa }^\dagger e^{i\epsilon_\kappa  \eta})\ ,\\
     c_j^\dagger(\eta)&= \frac{1}{\sqrt{L}}\sum_{\kappa \in \mathcal{K}} e^{i\kappa j}(v_\kappa  \gamma_\kappa  e^{-i\epsilon_\kappa  \eta} + u_\kappa  \gamma_{-\kappa }^\dagger e^{i\epsilon_\kappa  \eta})\ ,
    \end{split}
    \label{csol}
\end{equation}
where the domain of the wavenumber $\kappa $ is defined by
\begin{equation}
\mathcal{K}=\left\{\frac{2\pi}{L} \left(n-\frac{1}{2}\right)  \bigg| n=-\frac{L}{2}+1,\ldots,\frac{L}{2}\right\}\ ,
\label{Kdef_disc}
\end{equation}
and the time-independent oprators $\gamma_\kappa , \gamma^\dagger_\kappa $ satisfy  
\begin{equation}
    \{\gamma_\kappa ,\gamma^\dagger_{\kappa '}\}=\delta_{\kappa \kappa '}\ ,\quad \{\gamma_\kappa ,\gamma_{\kappa '}\}=\{\gamma^\dagger_\kappa ,\gamma^\dagger_{\kappa '}\}=0\ .
\end{equation}
See appnedix~\ref{TFI} for detailed calculations.
We have also introduced 
\begin{equation}
\begin{split}
    &z_\kappa =\frac{1-\cos \kappa}{\varepsilon}\ ,\quad y_\kappa =\frac{\sin \kappa }{\varepsilon}\  ,\quad \epsilon_\kappa =\sqrt{z_\kappa ^2+y_\kappa ^2}=\frac{2}{\varepsilon}|\sin\frac{\kappa}{2}|\ ,\\
    &\pmat{u_\kappa  \\ v_\kappa } = \frac{1}{\sqrt{2\epsilon_\kappa (\epsilon_\kappa +z_\kappa )}}\pmat{\epsilon_\kappa +z_\kappa  \\ iy_\kappa }\ .
\end{split}
\label{zyuvdef}
\end{equation}
Substituting Eq.~(\ref{csol}) into the Hamiltonian~(\ref{Hforc}), we have
\begin{equation}
    H=\sum_{\kappa \in \mathcal{K}} \epsilon_\kappa  \gamma_\kappa ^\dagger \gamma_{\kappa } + E_0\ ,
\end{equation}
where $E_0=-\sum_{\kappa \in \mathcal{K}} \epsilon_\kappa /2 = -\frac{L}{2\pi}\csc(\pi/2L)$.
As in the case of the continuum theory, we define the Bunch-Davies vacuum as
\begin{equation}
    \gamma_\kappa  |\Omega\rangle =0\quad (\forall \kappa \in \mathcal{K})\ .
    \label{BDdefspin}
\end{equation}

\subsection{Detector response in the spin model}
\label{subsec:detector_spin}

In terms of the spin system, the scalar operator that couples to the detector is given by
\begin{equation}
    \Phi(\eta,x_j)=H\cos\eta \nord{\Psi^\dagger(\eta, x_j) \Psi(\eta, x_j)}=\frac{H\cos\eta}{\varepsilon} \nord{c_j^\dagger(\eta) c_j(\eta)}\ .
    \label{Phiandc}
\end{equation}
From Eqs.~(\ref{csol}) and (\ref{BDdefspin}), 
the Wightman function for this scalar operator in the spin system is computed as 
\begin{multline}
    \langle \Omega | \Phi(\eta,x_j) \Phi(\eta',x_{j'})|\Omega \rangle \\
    =-\frac{H^2\cos\eta\cos \eta'}{4\pi^2}\sum_{\kappa,\kappa '\in \mathcal{K}}u_{\kappa '}v_{\kappa }(u_{\kappa '}v_{\kappa }-u_{\kappa }v_{\kappa '})
    e^{-i(\epsilon_{\kappa }+\epsilon_{\kappa '})\Delta \eta-i(\kappa +\kappa ')\Delta j}
    ,
\end{multline}
where $\Delta\eta=\eta-\eta'$ and $\Delta j=j-j'$. (See appendix~\ref{TFI} for detailed calculations.)
The detector response is given by
\begin{equation}
    \mathcal{F}(\omega)=\int^{\tau_1}_{\tau_0} d\tau'\int^{\tau_1}_{\tau_0} d\tau e^{-i\omega (\tau-\tau')} h(\tau)h(\tau')\langle \Phi(\eta(\tau),0) \Phi(\eta(\tau'),0)\rangle \ .
\end{equation}
The relation between the conformal time $\eta$ and the detector's proper time $\tau$ is the same as Eq.~(\ref{taueta}). 
We numerically compute it taking the contour~(\ref{contourc123}).

\begin{figure}[t]
  \centering
\subfigure[$H\mathcal{T}=3$]
 {\includegraphics[scale=0.45]{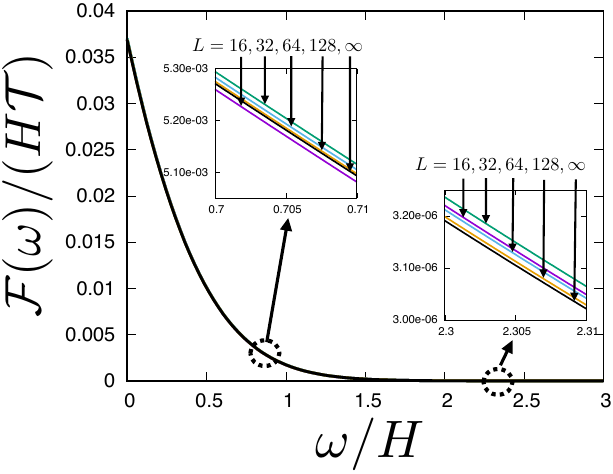}\label{dT5}
  }
  \subfigure[$H\mathcal{T}=5$]
 {\includegraphics[scale=0.45]{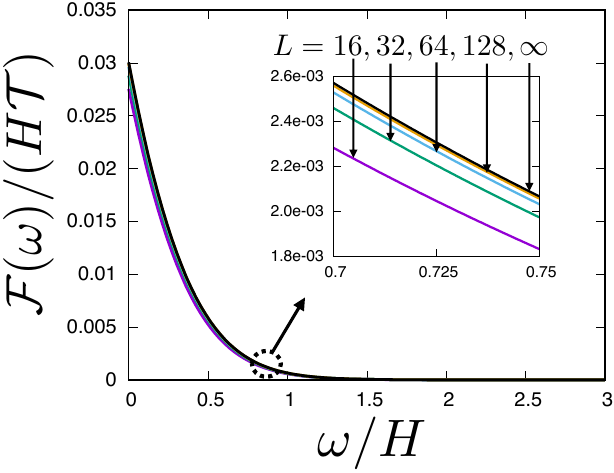}\label{dT10}
  }
   \subfigure[$H\mathcal{T}=10$]
 {\includegraphics[scale=0.45]{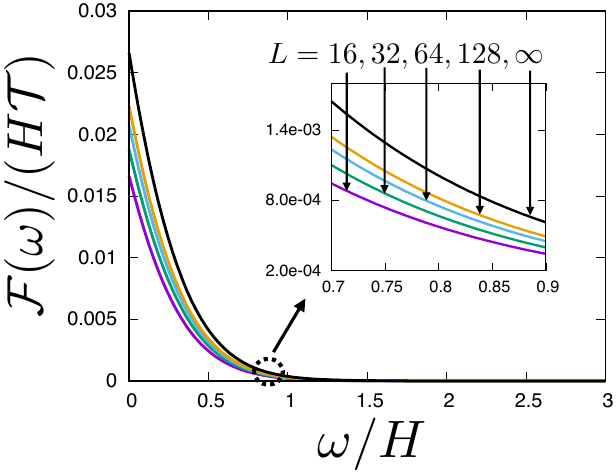}\label{dT20}
  }
 \caption{Detector response computed by the spin model for $H\mathcal{T}=3,5,10$. In each panel, the number of spins are varied as $L=16,32,64,128$. 
}
 \label{F_spin}
\end{figure}

Figure~\ref{F_spin} shows the detector response computed using the spin model with $L=16,32,64,128$. The black curves are results for QFT in the continuum limit ($L=\infty$). The effective measurement time is varied as (a) $H\mathcal{T}=3$, (b) $H\mathcal{T}=5$ and (c) $H\mathcal{T}=10$. 
In panel (a),
the detector response takes almost identical values for $L=16,32,64,128,\infty$, making it difficult to distinguish each curve in the overall plot. For clarity, the insets with magnified views are provided. We observe convergence of the detector response to that of QFT as $L$ increases. Although there are differences in the rate of convergence, all panels (a), (b), and (c) exhibit convergence to the results of QFT. We will discuss the origin of the differences in the rate of convergence in the next subsection.

\begin{figure}[t]
  \centering
\subfigure[$H\mathcal{T}=3$]
 {\includegraphics[scale=0.45]{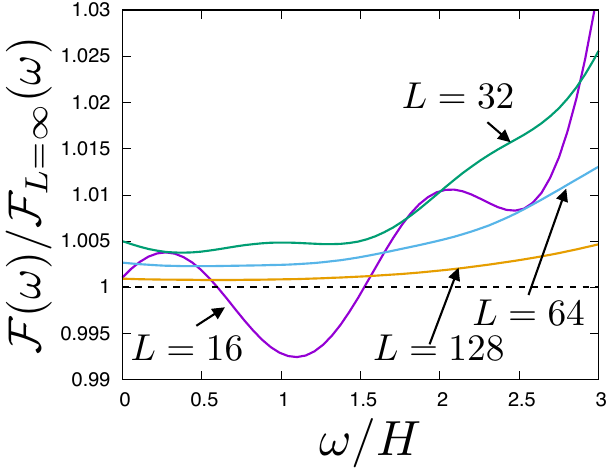}\label{dT5_ratio}
  }
  \subfigure[$H\mathcal{T}=5$]
 {\includegraphics[scale=0.45]{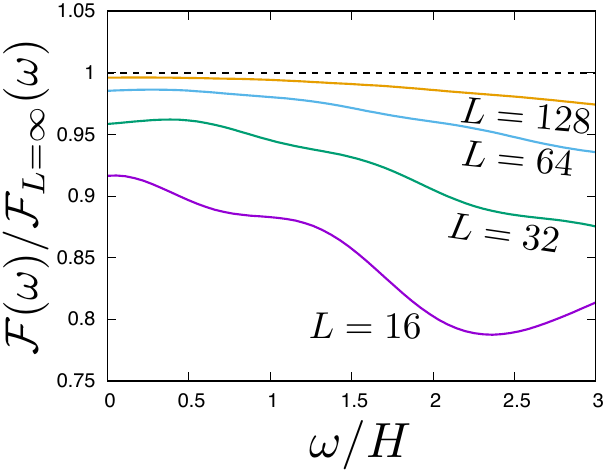}\label{dT10_ratio}
  }
   \subfigure[$H\mathcal{T}=10$]
 {\includegraphics[scale=0.45]{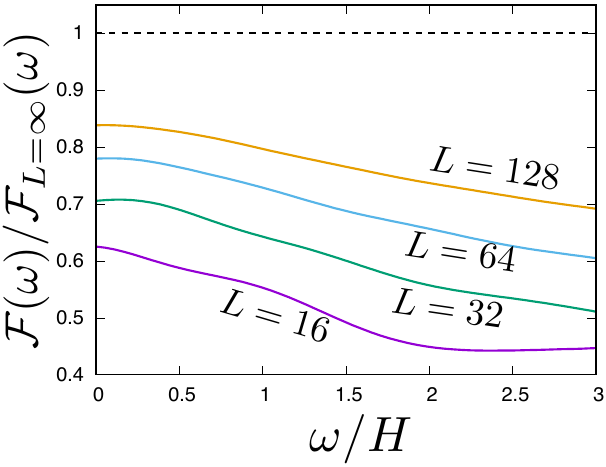}\label{dT20_ratio}
  }
 \caption{The ratio of the detector response in the spin model to that of QFT for $H\mathcal{T}=3,5,10$. In each panel, the number of spins are varied as $L=16,32,64,128$. 
}
 \label{F_spin_ratio}
\end{figure}

Figure~\ref{F_spin_ratio} shows the ratio of the detector response in the spin model to that of QFT: 
\begin{equation}
    \frac{\mathcal{F}(\omega)}{\mathcal{F}_{L=\infty}(\omega)}\ .
\end{equation}
This figure illustrates more clearly how the results of the spin system converge to those of QFT as $L$ increases. Each panel shows that even in the large $\omega$ regime, the detector responses for the spin model and the QFT differ by at most a factor of a few. 
This indicates that the detector response behaves as $\mathcal{F}(\omega) \sim e^{-\beta\omega}$ even for finite $L$, allowing us to measure the temperature of de Sitter spacetime using the spin system.

\begin{figure}[t]
  \centering
\subfigure[$H\mathcal{T}=3$]
 {\includegraphics[scale=0.45]{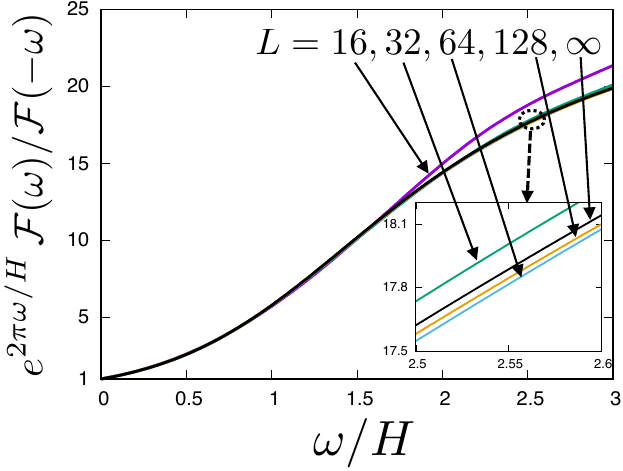}\label{KMSdT5}
  }
  \subfigure[$H\mathcal{T}=5$]
 {\includegraphics[scale=0.45]{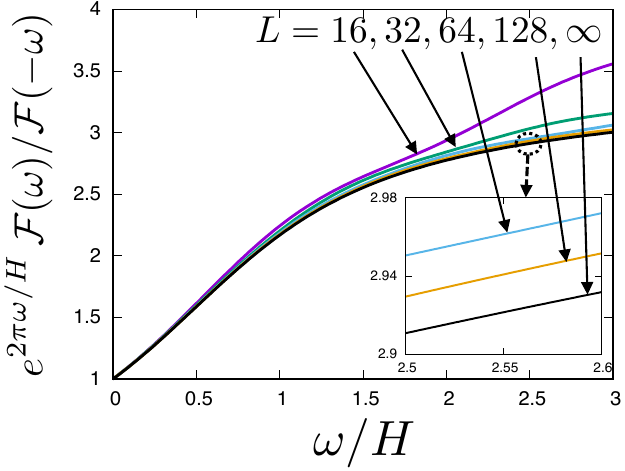}\label{KMSdT10}
  }
   \subfigure[$H\mathcal{T}=10$]
 {\includegraphics[scale=0.45]{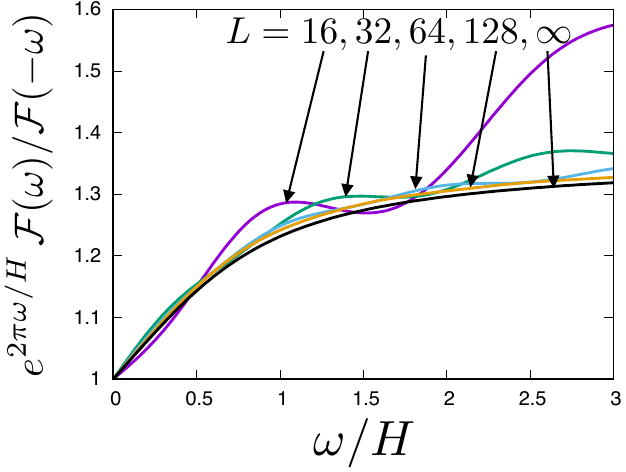}\label{KMSdT20}
  }
 \caption{The ratio of the detector response in the spin model to that of QFT for $H\mathcal{T}=3,5,10$. In each panel, the number of spins are varied as $L=16,32,64,128$. 
}
 \label{KMS}
\end{figure}

So far, although we have focused only on the frequency with $\omega\geq 0$, it is also possible to calculate the detector response in the $\omega<0$ region. As discussed in section \ref{UDdet}, by taking a sufficiently long measurement time and evaluating the detector response over $\omega \in \bm{R}$, we can obtain complete information about the Wightman functions $G^\pm(\omega)$ from the detector response. Here, using the results of the detector response for negative frequencies, we evaluate
\begin{equation}
    e^{2\pi \omega/H} \frac{\mathcal{F}(\omega)}{\mathcal{F}(-\omega)}\ .
\end{equation}
If this quantity becomes one, it is equivalent to satisfying the KMS relation~(\ref{KMSforF}). In a thermal Minkowski spacetime, when the measurement time is sufficiently long, this value becomes exactly equal to one. Figure~\ref{KMS} shows the above quantity for $H\mathcal{T}=3,5,10$. 
Typically, $\mathcal{F}(\omega)/\mathcal{F}(-\omega)$ behaves as $\sim e^{-2\pi\omega/H}$, but its coefficient deviates from 1. However, as expected, it can be seen that the coefficient approaches 1 as the measurement time increases. This result suggests that the KMS relation is well satisfied when the measurement time and the number of sites are taken to be sufficiently large. (As described in the next subsection, it is necessary to choose the number of sites such that $L \gg e^{H\mathcal{T}}$ is satisfied, depending on the measurement time.)

\subsection{Trans-Plankian problem}

We find that the convergence to the results in the QFT appears to be slower as the value of $H\mathcal{T}$ increases. 
This indicates that, when the detector interacts with the spin system for a long period, the spin system is no longer a good approximation to the QFT. This phenomenon is analogous to the Trans-Planckian problem in cosmology~\cite{Brandenberger:1999sw}. 
We simulate the QFT in the de Sitter spacetime using a finite number of spins. The coordinate spacing between spins is given by $\varepsilon = 2\pi / L$, which provides the minimum length of the system. (This is analogous to the Planck length in our universe.) 
In terms of the wave number $k$, this gives a UV cutoff $k_{\mathrm{UV}} \simeq 2\pi / \varepsilon = L$. As in Eq.~(\ref{Minksol}), the mode function with wave number $k$ behaves as $\sim \exp(\pm i |k| \eta + i k x)$. In cosmological time $t$, this can be expressed as $\sim \exp(\pm i |k| (\pi/2 - 2 e^{-Ht}) + i k x)$ for $t\gg 1/H$.
It follows that the mode function with wave number $k$ does not oscillate for $|k| e^{-Ht} \lesssim 1$. (In cosmology, this phenomenon is known as freeze-out.) In other words, at a given time $t$, only the modes with $|k| \gtrsim e^{Ht}$ are dynamical, and the QFT is described by these modes. Thus, in the spin model, it is required that $k_{\mathrm{UV}} \simeq L \gg e^{Ht}$ for it to accurately describe the QFT.

At a cosmological time $t$, a physical length scale should be less than the Hubble scale $H^{-1}$, which means that a proper distance satisfies $e^{Ht} \Delta x/H \ll 1/H$. On the other hand, a coordinate distance should be $\Delta x \gtrsim \epsilon\sim 1/L$ in the spin model.
Thus, at sufficiently late time, $t \gtrsim H^{-1} \log L$, the Planck scale will be expanded beyond the Hubble scale.
If the measurement time $\mathcal{T}$ is too large, it is inevitable that it will be affected by this late-time behavior.

\begin{figure}[t]
  \centering
\subfigure[$H\mathcal{T}=2$]
 {\includegraphics[scale=0.45]{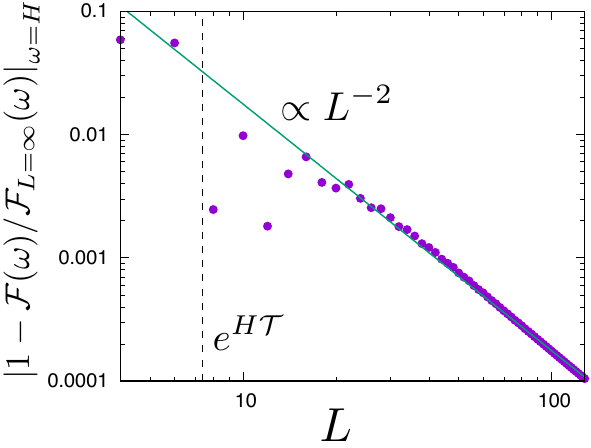}\label{cdt2}
  }
  \subfigure[$H\mathcal{T}=2.5$]
 {\includegraphics[scale=0.45]{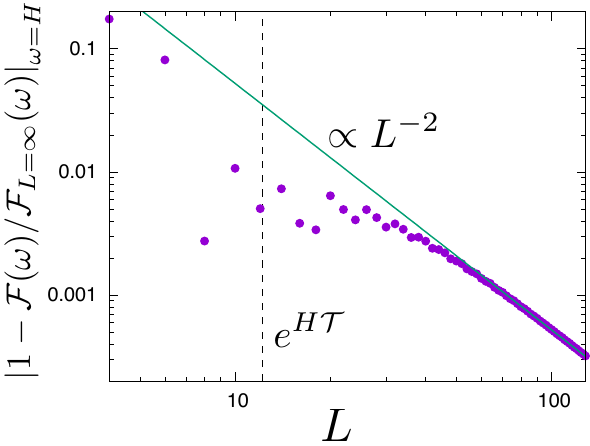}\label{cdt25}
  }
  \subfigure[$H\mathcal{T}=3$]
 {\includegraphics[scale=0.45]{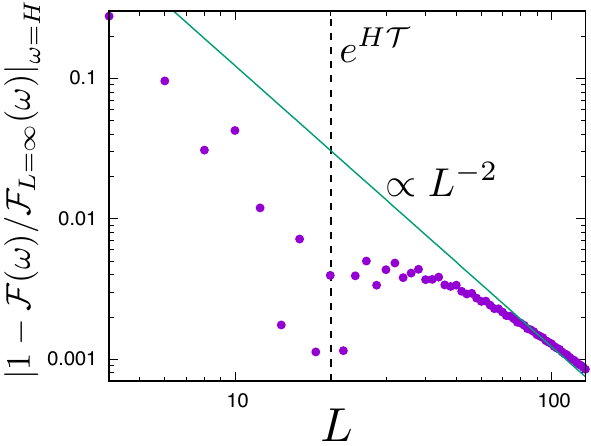}\label{cdt3}
  }
  \subfigure[$H\mathcal{T}=3.5$]
 {\includegraphics[scale=0.45]{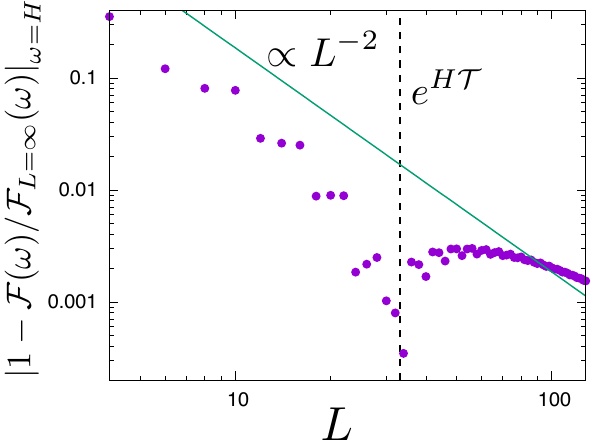}\label{cdt35}
  }
  \subfigure[$H\mathcal{T}=4$]
 {\includegraphics[scale=0.45]{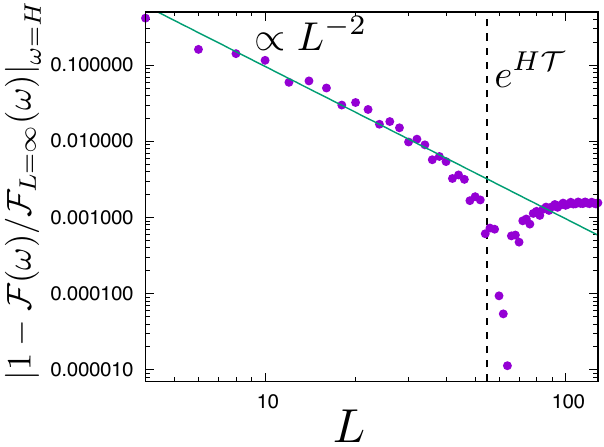}\label{cdt4}
  }
  \caption{
 The convergence of the detector response with respect to the number of sites $L$. Panels (a)–(e) correspond to different effective measurement times, $H\mathcal{T} = 2, 2.5, 3, 3.5, 4$. The vertical dashed lines indicate $e^{H\mathcal{T}}$. The green lines represent $\propto L^{-2}$ fitted using data for $L = 64,\ldots,128$.
}
 \label{conv}
\end{figure}

Figure~\ref{conv} illustrates the convergence of the detector response with respect to $L=4, 6, \ldots, 128$. 
The vertical axis is $|1-\mathcal{F}(\omega)/\mathcal{F}_{L=\infty}(\omega)|$ evaluated at $\omega=H$. 
Panels (a)-(e) correspond to the results for effective measurement time $H\mathcal{T}=2, 2.5, 3, 3.5, 4$, respectively. In each panel, the position of $e^{H\mathcal{T}}$ is indicated by the vertical dashed line. 
Roughly speaking, the left side of the line behaves 
``discrete system-like,''
while the right side behaves 
``continuous system-like.''
To the left of that line, the spin system fails to approximate QFT well due to the Trans-Planckian problem. The green line represents  the curve of $\propto L^{-2}$ fitted using data for $L=64,\ldots, 128$. In the region where $L\gg e^{H\mathcal{T}}$, especially for panels (a)-(c), good convergence obeying a power law $\sim L^{-2}$ can be observed. (For Panels (d) and (e), since the number of data points satisfying $L \gg e^{H\mathcal{T}}$ is limited, the power law is not clearly observed.)

\section{Implementation of Unruh-DeWitt detector in spin systems: Hamiltonian for quantum simulation}
\label{single_spin}

As a concrete example of a detector in the spin system, we consider a single spin whose Hamiltonian is given by
\begin{equation}
    \hat{H}_\textrm{d}=\frac{\omega}{2}\sigma^z\ .
    \label{Hdsinglespin}
\end{equation}
Its eigenstates are denoted by $\dw$ and $\up$. The eigenvalues are $-\omega/2$ and $\omega/2$, respectively. We assume that the initial state of the detector is $\dw$. We will consider the transition probability of $\dw \to \up$ due to the interaction between the detector and the quantum field.
As the detector's monopole moment operator, we adopt
\begin{equation}
    m=\sigma^x\ .
\end{equation}
From Eq.~(\ref{PE}), the transition probability from $\dw$ to $\up$ is 
\begin{equation}
P(\dw \to \up) 
    =\lambda^2 |\upbra  \sigma^x  \dw|^2 \mathcal{F}(\omega)
    =\lambda^2 \mathcal{F}(\omega)\ .
\end{equation}
It follows that we can measure the detector response through the transition probability of the detector as
\begin{equation}
    \mathcal{F}(\omega) = \frac{1}{\lambda^2} P(\dw \to \up) \ .
\end{equation}
By repeating the experiment while sweeping frequencies, we can obtain the detector response $\mathcal{F}$ as a function of $\omega$.

When the number of sites $L$ is sufficiently large ($L\gg e^{Ht}$), 
the spin system~(\ref{Hspin}) provides an approximation to QFT of Majorana fermions~(\ref{Hds}). 
The scalar operator $\Phi$ in QFT is written by operators in the spin system as in Eq.~(\ref{Phiandc}).
Using the spin operator, it is rewritten as 
\begin{equation}
    \Phi(\eta,x_j) = -\frac{H\cos\eta}{2\varepsilon}(\sigma^z_j(\eta) - \langle \sigma^z\rangle) \ ,
\end{equation}
where $\langle \sigma^z\rangle\equiv \langle \Omega |\sigma_j^z(\eta) |\Omega\rangle$. 
From the solution of the Heisenberg equation~(\ref{csol}), this is explicitly written as
\begin{equation}
\langle \sigma^z\rangle = 1-\frac{2}{L}\sum_{\kappa\in\mathcal{K}} |v_k|^2\ .
\end{equation} 
This does not depend on space $j$ and time $\eta$. 

When we take the single spin introduced in Eq.~(\ref{Hdsinglespin}) as the detector, 
the total Hamiltonian of QFT and detector~(\ref{Htot}) is given by
\begin{equation}
    \hat{H}_\textrm{tot}(\eta)=\hat{H}_\textrm{Spin}\otimes 1 + 1\otimes \frac{1}{H\cos\eta} \frac{\omega}{2}\sigma^z  + \frac{\lambda h(\tau)}{2\varepsilon}(\sigma^z_1(\eta) - \langle \sigma^z\rangle)\otimes \sigma^x .
\end{equation}
Just by solving this system and computing the probability of $\dw \to \up$, we have the detector response. 
The above Hamiltonian describes the time evolution in terms of the conformal time $\eta$. The proper time $\tau$ in the argument of the window function $h(\tau)$ is determined from $\eta$ via Eq.~(\ref{taueta}). 

\section{Conclusion}
In this study, we have developed a novel framework to probe the thermal properties of quantum field theory (QFT) in an inflationary universe by leveraging spin systems as a simulable analog. Building on our previous work, which established a mapping between the QFT of Majorana fermions in arbitrary two-dimensional spacetimes and spin systems~\cite{Kinoshita:2024ahu}, we extended this approach to the context of an inflationary universe, modeled as the de Sitter spacetime. By coupling an Unruh-DeWitt detector to both the continuum QFT and its discrete spin system counterpart, we investigated the detector's excitation probability as a means to characterize the thermal nature of the quantum field.

Our analysis in Section~\ref{sec:UDdeSitter} demonstrated that, in the continuum limit, a static observer in de Sitter spacetime perceives a thermal bath with a temperature \( H/(2\pi) \), consistent with the Gibbons-Hawking effect. The detector response, derived from the Wightman function of Majorana fermions, exhibits a Planckian distribution, confirming the thermal character of the inflationary universe. In Section~\ref{sec:spin_deSitter}, we translated this scenario into a spin system framework, specifically the transverse-field Ising model at its critical point, and showed that the detector response in the spin model converges to the QFT result as the number of spin sites \( L \) increases. This convergence, however, is modulated by the effective measurement time \( \mathcal{T} \) and the system size, with deviations arising due to the Trans-Planckian problem when \( L \ll e^{H \mathcal{T}} \). 

A key practical advancement of this work, detailed in Section~\ref{single_spin}, is the proposal of a concrete implementation of the Unruh-DeWitt detector within a spin system. By modeling the detector as a single spin with a Hamiltonian \( \hat{H}_{\mathrm{d}} = \frac{\omega}{2} \sigma^z \) and coupling it to the spin system via the operator \( \Phi(\eta, x_j) \), we derived a total Hamiltonian amenable to quantum simulation. The transition probability \( P(|\downarrow\rangle \to |\uparrow\rangle) \) directly yields the detector response \( \mathcal{F}(\omega) \), offering a measurable quantity that can be experimentally realized on quantum computing platforms. This approach bridges the gap between theoretical predictions and experimental verification, providing a pathway to test the thermal properties of an inflationary universe in a controlled laboratory setting.

Our findings suggest that spin systems, despite their discrete nature, can effectively approximate the thermal behavior of QFT in curved spacetimes, provided the system size and measurement time are appropriately tuned. The satisfaction of the Kubo-Martin-Schwinger (KMS) relation in the large-\( L \) and long-\( \mathcal{T} \) limits further validates the thermal interpretation of the detector response in the spin model. However, the Trans-Planckian problem highlights a limitation: as \( \mathcal{T} \) increases, an exponentially large number of spins is required to maintain fidelity with the continuum theory, posing a challenge for practical simulations.

Importantly, these findings not only deepen our understanding of the thermal nature of quantum fields in curved spacetimes, but also open the door to experimental exploration of these effects. In particular, our approach offers a practical means of accessing the effective temperature associated with cosmological horizons---an observable that has long been established in theory but has remained difficult to probe directly due to the extremely weak nature of quantum gravitational effects in our universe. By embedding an Unruh-DeWitt-like detector within a spin system and relating its excitation probability to measurable quantities, we provide a concrete and scalable framework for investigating horizon-induced thermal behavior in controllable quantum systems. With the ongoing development of programmable quantum simulation platforms~\cite{Daley2022-ok}---such as superconducting qubits~\cite{Devoret2013-bg,Wendin2017-io,Houck2012-hc,Kim2023-nz}, trapped ions~\cite{Barreiro2011-cp,Blatt2012-fa,Monroe2021-tn}, and Rydberg atom arrays~\cite{Weimer2010-fu,Henriet2020-yk}---these ideas are becoming increasingly feasible to realize in the laboratory. In this way, our work helps to connect theoretical predictions with experimental possibilities, and contributes to the growing effort to simulate quantum field theory in curved spacetimes using table-top quantum systems.

This work opens several avenues for future research. First, extending the spin model to include massive Majorana fermions or higher-dimensional spacetimes could provide deeper insights into the robustness of our approach across different physical regimes. Second, optimizing the quantum simulation protocol---perhaps by exploring alternative window functions or detector couplings---could mitigate finite-size effects and enhance convergence rates. Finally, experimental realization on near-term quantum hardware would serve as a critical test of our predictions, potentially shedding light on the quantum nature of inflationary cosmology. By establishing spin systems as a viable tool for simulating QFT in curved spacetimes, this study lays the groundwork for a new intersection of quantum information science and cosmology.

\begin{acknowledgments}
We would like to thank Tadashi Kadowaki for useful discussions and comments.
The work of K.\ M.\ was supported in part by JSPS KAKENHI Grant Nos.\ JP20K03976, JP21H05186 and JP22H01217.
The work of D.\ Y.\ was supported by JSPS KAKENHI Grant Nos.~21H05185, 23K25830, and 24K06890, and JST PRESTO Grant No.~JPMJPR245D. 
The work of R.\ Y.\ was supported by JSPS KAKENHI Nos.~19K14616, 20H01838, and 25K07156. 
\end{acknowledgments}

\appendix

\appendix

\section{Brief review of the detector response function}
\label{derivation_detecorres}

For the following discussion, it is convenient to adopt the detector's proper time as the time coordinate of the spacetime applying the coordinate transformation, $t = t(\tau)$. 
Then, the Hamiltonian~(\ref{Htot}) becomes
\begin{equation}
    \hat{H}_\textrm{tot}(\tau)=\hat{H}_\textrm{QFT}(\tau)\otimes 1 + 1 \otimes \hat{H}_\textrm{d}+ \lambda V(\tau)\ .
\end{equation}
We will denote the unperturbed Hamiltonian as
\begin{equation}
    \hat{H}_0(\tau)\equiv \hat{H}_\textrm{tot}(\tau)|_{\lambda=0}=\hat{H}_\textrm{QFT}(\tau)\otimes 1 + 1 \otimes \hat{H}_\textrm{d} \ .
\end{equation}

Consider a quantum state for the total system $|\psi(\tau)\rangle \in \mathcal{H}_\textrm{QFT}\otimes \mathcal{H}_\textrm{d}$, and its time evolution. We introduce a time evolution operator for the unperturbed Hamiltonian as
\begin{equation}
\begin{split}
    &U_0(\tau,\tau_0) = U_\text{QFT}(\tau,\tau_0) \otimes U_\text{d}(\tau,\tau_0)\ ,\\
    &U_\text{QFT}(\tau,\tau_0)=T \{e^{-i\int_{\tau_0}^\tau \hat{H}_\textrm{QFT}(\tau') d\tau'} \}\ ,\quad 
    U_\text{d}(\tau,\tau_0)=e^{-i\hat{H}_\text{d} (\tau-\tau_0)}\ .
\end{split}
\end{equation}
Defining a state in the interaction picture as $|\psi(\tau)\rangle_I = U_0^\dagger(\tau,\tau_0)|\psi(\tau)\rangle$, we can write the Schr\"odinger equation as
\begin{equation}
    i\frac{d}{d\tau} |\psi(\tau)\rangle_I = \lambda V_I(\tau)  |\psi(\tau)\rangle_I\ ,
\end{equation}
where $V_I(\tau)=U_0^\dagger (\tau,\tau_0) V(\tau) U_0 (\tau,\tau_0)$. 
Since $U_0 (\tau,\tau_0)$ is a separable operator in $\mathcal{H}_\textrm{QFT}\otimes \mathcal{H}_\textrm{d}$, we have
\begin{equation}
V_I(\tau) = -h(\tau) \, \Phi(\tau,\bm{x}(\tau))\otimes m(\tau) \ ,
\end{equation}
where 
\begin{equation}
\begin{split}
    &\Phi(\tau,x(\tau))= U^\dagger_\text{QFT}(\tau,\tau_0) \Phi(\bm{x}(\tau)) U_\text{QFT}(\tau,\tau_0) ,\\
    &m(\tau)=U^\dagger_\text{d}(\tau,\tau_0)\, m\, U_\text{d}(\tau,\tau_0)\ .
    \end{split}
\end{equation}
They are nothing but Heisenberg operators without interaction. 
We expand the state in terms of the coupling constant $\lambda$ into 
$|\psi(\tau)\rangle_I = |\psi(\tau_0)\rangle + \lambda |\delta \psi(\tau)\rangle_I + \cdots$.
Then in the first order of $\lambda$, we have
\begin{equation}
    |\delta \psi(\tau)\rangle_I = -i \int^\tau_{\tau_0}d\tau' V_I(\tau') |\psi(\tau_0)\rangle\ .
\end{equation}

We assume that the initial state is the ground state in $\mathcal{H}_\textrm{QFT}\otimes \mathcal{H}_\textrm{d}$: 
\begin{equation}
    |\psi(\tau_0)\rangle = |\Omega\rangle \otimes |E_0\rangle\ .
\end{equation}
For $\tau>\tau_1$, the transition amplitude to $|\varphi\rangle \otimes |E\rangle$ ($E\neq E_0$) is given by
\begin{multline}
    \mathcal{M}(E_0\to E; \Omega \to \varphi)=\lambda (\langle \varphi| \otimes \langle E|) |\delta \psi(\tau_0)\rangle\\
    =i\lambda \int^{\tau_1}_{\tau_0}d\tau' h(\tau')
    \langle \varphi|\Phi(\tau',\bm{x}(\tau'))|\Omega\rangle 
    \langle E | m(\tau') |E_0\rangle \ .
\end{multline}
Thus, the probability that the energy of the detector becomes $E$ is
\begin{equation}
\begin{split}
    &P(E_0\to E) = \sum_{\varphi} |\mathcal{M}(E_0\to E; \Omega \to \varphi)|^2\\
    &=\lambda^2 \int^{\tau_1}_{\tau_0}d\tau' \int^{\tau_1}_{\tau_0}d\tau h(\tau)h(\tau')\\
    &\hspace{3cm}\times \langle \Omega |\Phi(\tau,\bm{x}(\tau)) \Phi(\tau',\bm{x}(\tau'))|\Omega\rangle 
    \langle E_0 | m(\tau) |E\rangle  \langle E | m(\tau') |E_0\rangle \ ,
    \end{split}
\end{equation}
where we used complete relation $\sum_\varphi|\varphi\rangle \langle \varphi|=1$. 
Because of $\langle E | m(\tau') |E_0\rangle=e^{i(E-E_0)\tau} \langle E | m |E_0\rangle$, 
the above expression becomes
\begin{equation}
    P(E_0\to E) 
    =\lambda^2 |\langle E | m  |E_0\rangle|^2 \mathcal{F}(E-E_0) \ ,
    \label{PE2}
\end{equation}
where 
\begin{equation}
    \mathcal{F}(\omega)=\int^{\tau_1}_{\tau_0}d\tau' \int^{\tau_1}_{\tau_0}d\tau h(\tau)h(\tau')
    e^{-i\omega (\tau-\tau')}
    \langle \Omega |\Phi(\tau,\bm{x}(\tau)) \Phi(\tau',\bm{x}(\tau'))|\Omega\rangle 
    \ .
    \label{Fdef0}
\end{equation}
The function $\mathcal{F}(\omega)$ is called detector response function and does not depend on the detail of the detector.

In the original coordinate system $x = (t, \bm{x})$, we simply have $\Phi(t, \bm{x}) = \Phi(\tau, \bm{x})$, as we assumed that $\Phi$ is a scalar operator. Thus, the detector response is given by
\begin{equation}
    \mathcal{F}(\omega)=\int^{\tau_1}_{\tau_0}d\tau' \int^{\tau_1}_{\tau_0}d\tau h(\tau)h(\tau')
    e^{-i\omega (\tau-\tau')} G^+(x(\tau),x(\tau'))
    \ ,
\end{equation}
where we have introduced the Wightman function as
\begin{equation}
    G^+(x,x')=\langle \Omega |\Phi(x)\Phi(x')|\Omega\rangle \ .
\end{equation}
In this sense, the Fourier transform of the Wightman function, multiplied by a window function, is the detector's response.

\section{Wightman functions for a scalar operator of Majorana fermions in flat spacetime}

\label{WightmanMink}

\subsection{Solution of the Heisenberg equation}

The Heisenberg equation for $\Psi$ is given by
\begin{equation}
    \dot{\Psi}-i\Psi^\dagger{}'=0\ ,\quad \dot{\Psi}^\dagger+i\Psi'=0\ .
\end{equation}
We apply the Fourier transform as
\begin{equation}
    \Psi(\eta,x)=\frac{1}{\sqrt{\ell}}\sum_{k\in K} e^{ikx} \Psi_k(t)\ ,\quad 
    \Psi^\dagger (\eta,x)=\frac{1}{\sqrt{\ell}}\sum_{k\in K} e^{ikx} \Psi^\dagger_{-k}(t)\ ,
\end{equation}
where, from the anti-periodic boundary condition, the domain of the wave number is
\begin{equation}
K=\left\{\frac{2\pi}{\ell}\left(n-\frac{1}{2}\right)\bigg|n\in \bm{Z}\right\}\ .
\label{Kdef_cont2}
\end{equation}
Then, the Heisenberg equations are rewritten as 
\begin{equation}
\left[\frac{d}{dt} + k\pmat{0 & 1\\-1 & 0}\right]\pmat{\Psi_k(t)\\\Psi_{-k}^\dagger(t)} = 0\ .
\end{equation}
Decomposing the time-dependence of variables into the Fourier modes as $(\Psi_k, \Psi_{-k}^\dagger) \propto e^{-i\omega t}$, we obtain eigen-frequencies and their eigen-vectors as
\begin{equation}
\begin{split}
    \omega =|k|\ ,\quad \frac{1}{\sqrt{2}}\pmat{1 \\ i\mathrm{sgn}(k)}\ ,\\
    \omega =-|k|\ ,\quad \frac{1}{\sqrt{2}}\pmat{i\mathrm{sgn}(k) \\ 1}\ .
\end{split}
\end{equation}
Thus, the general solution of the Heisenberg equation in the momentum space is given by
\begin{equation}
    \pmat{\Psi_k(t)\\\Psi_{-k}^\dagger(t)} = \frac{1}{\sqrt{2}}\left[\gamma_k e^{-i|k|t} \pmat{1 \\ i\mathrm{sgn}(k)} + \gamma_{-k}^\dagger e^{i|k|t} \pmat{i\mathrm{sgn}(k) \\ 1}\right]\ .
\end{equation}
In the position space, we obtain
\begin{equation}
\begin{split}
    &\Psi(t,x)=\frac{1}{\sqrt{2\ell}}\sum_{k\in K} e^{ikx} (\gamma_k e^{-i|k|t}+i\mathrm{sgn}(k) \gamma_{-k}^\dagger e^{i|k|t})\ ,\\
    &\Psi^\dagger(t,x)=\frac{1}{\sqrt{2\ell}}\sum_{k\in K} e^{ikx} (i\mathrm{sgn}(k) \gamma_k e^{-i|k|t}+\gamma_{-k}^\dagger e^{i|k|t})\ .
\end{split}
\end{equation}
Defining the set of the positive wavenumber as $K^+=\{k>0, k\in K\}$, above expressions are rewritten as
\begin{equation}
\begin{split}
    &\Psi(t,x)=\frac{1}{\sqrt{2\ell}}\sum_{k\in K^+} (\gamma_k e^{-iku}+i\gamma_{-k}^\dagger e^{ikv}
    +\gamma_{-k} e^{-ikv}-i\gamma_{k}^\dagger e^{iku})\ ,\\
    &\Psi^\dagger(t,x)=\frac{1}{\sqrt{2\ell}}\sum_{k\in K^+} (i\gamma_{k} e^{-iku}+\gamma_{-k}^\dagger e^{ikv}-i\gamma_{-k} e^{-ikv}
    +\gamma_k^\dagger e^{iku})\ ,
\end{split}
\label{Minksol2}
\end{equation}
where $u=t-x$ and $v=t+x$.
Using the canonical anti-commutation relation, we obtain 
\begin{equation}
    \{\gamma_k,\gamma_{k'}^\dagger\}=\delta_{k,k'}\ ,\quad 
    \{\gamma_k,\gamma_{k'}\}=\{\gamma_k^\dagger,\gamma_{k'}^\dagger\}=0\ .
    \label{gammacomu2}
\end{equation}
Substituting Eq.~(\ref{Minksol}) into Eq.~(\ref{Hmink}), we can write the Hamiltonian of the Majorana fermion as 
\begin{equation}
    H=\sum_{k\in K} |k| N_k  + E_0\ ,\quad N_k=\gamma_k^\dagger \gamma_k\ ,
    \label{Hk2}
\end{equation}
where $E_0=-\sum_k |k|/2 = -\infty$ is the vacuum energy. Therefore, the vacuum state $|\Omega\rangle$ is defined by 
\begin{equation}
    \gamma_k |\Omega\rangle =0 \ ,\quad (\forall k\in K)\ .
    \label{vacdef2}
\end{equation}
Excited states can be constructed by multiplying creation operators to $|\Omega\rangle$. The Fock space is spanned by
\begin{equation}
    [\cdots (\gamma_{k_{-2}}^\dagger)^{m_{-2}}(\gamma_{k_{-1}}^\dagger)^{m_{-1}}(\gamma_{k_{0}}^\dagger)^{m_{0}}
    (\gamma_{k_{1}}^\dagger)^{m_{1}}(\gamma_{k_{2}}^\dagger)^{m_{2}}\cdots ]|\Omega\rangle \ ,
    \label{exstate}
\end{equation}
where $k_n = 2\pi (n-1/2) /\ell$ and $m_i = 0$ or $1$ ($i\in \bm{Z}$).

\subsection{Wightman function for the vacuum state}

We define the Hermitian scalar operator $\Phi$ as in Eq.~(\ref{Phidef}). 
The normal ordering is denoted by $:\mathcal{O}:$. By the normal ordering, all creation operators $\gamma_k^\dagger$ are to the left of all annihilation operators $\gamma_k$ in the operator $\mathcal{O}$, where the spin statistics are taken into account when exchanging the operators. For example, $:\gamma_k \gamma_{k'}^\dagger:=-\gamma_{k'}^\dagger \gamma_k$. We will compute the Wightman function for the operator $\Phi$:
\begin{equation}
    G^+(t,x,t',x')=\langle \Omega | \Phi(t,x) \Phi(t',x') |\Omega\rangle \ .
\end{equation}
Using the solution of the Heisenberg equation~(\ref{Minksol2}), we have
\begin{multline}
    \ell \Phi(t,x)=\sum_{k,k'\in K^+}(
    \gamma_{-k'}^\dagger \gamma_k e^{-iku+ik'v}
    +i\gamma_k \gamma_{-k'} e^{-iku-ik'v}\\
    -i\gamma_{-k'}^\dagger \gamma_k^\dagger e^{iku+ik'v}
    +\gamma_k^\dagger \gamma_{-k'} e^{iku-ik'v})\ .
    \label{Phi}
\end{multline}
From Eq.~(\ref{vacdef2}), the Wightman function becomes
\begin{equation}
\begin{split}
     G^+(t,x,t',x')&=\frac{1}{\ell^2}\sum_{k,k',p,p'\in K^+}e^{-iku-ik'v+ipu'+ip'v'} \langle \Omega |  \gamma_k \gamma_{-k'} \gamma_{-p'}^\dagger \gamma_p^\dagger |\Omega\rangle \\
     &=\frac{1}{\ell^2}\sum_{k,k'\in K^+}e^{-ik(u-u')-ik'(v-v')}\\
     &=g(u-u')g(v-v')\ ,
     \end{split}
\end{equation}
where we define
\begin{equation}
    g(u)\equiv \frac{1}{\ell} \sum_{k\in K^+} e^{-ik(u-i\epsilon)}=\frac{1}{2i\ell \sin[\frac{\pi}{\ell}(u-i\epsilon)]}\ ,
\end{equation}
where we infinitesimally shifted the coordinate $u$ on the complex plane to make the geometric series converge.
In the limit of $\ell\to \infty$, the function $g$ and corresponding Wightman function are given by
\begin{equation}
\begin{split}
    &g(u)|_{\ell\to\infty}=\frac{1}{2\pi i(u-i\epsilon)}\ ,\\  
    &G^+(t,x,t',x')|_{\ell\to\infty}=-\frac{1}{4\pi^2(u-u'-i\epsilon)(v-v'-i\epsilon)}\ .
    \end{split}
\end{equation}
This is consistent with the prediction of conformal field theory. We have determined the prefactor of the Wightman function by its direct calculation.

\subsection{Wightman function for the thermal state}

Here, we compute the thermal Wightman function for the operator $\Phi$, which is defined as
\begin{equation}
    G^+_\beta(t,x,t',x')=\langle  \Phi(t,x) \Phi(t',x') \rangle_\beta  \ .
    \label{thermalWight}
\end{equation}
We substitute Eq.~(\ref{Phi}) into the above equation. Then, we need to compute the thermal expectation values of the composite operators of $\gamma$ and $\gamma^\dagger$. For these computations, it is convenient to use the following formulas:
\begin{equation}
    e^{-\beta H}\gamma_k^\dagger = \gamma_k^\dagger e^{-\beta (H+|k|)}\ ,\quad 
    e^{-\beta H}\gamma_k = \gamma_k e^{-\beta (H-|k|)}\ .
    \label{gammaformula}
\end{equation}
These relations can be verified by multiplying both sides of Eq.~(\ref{exstate}).
Using the above equations and the cyclic property of the trace, we can also prove
\begin{equation}
    \langle \gamma_k^\dagger \mathcal{O} \rangle_\beta = e^{-\beta|k|} \langle \mathcal{O} \gamma_k^\dagger \rangle_\beta\ .
    \label{gammaformula2}
\end{equation}
for any operator $\mathcal{O}$.

As an exercise, let us consider $\langle \gamma_k^\dagger \gamma_p\rangle_\beta$. 
It is computed as
\begin{equation}
    \langle \gamma_k^\dagger \gamma_p\rangle_\beta 
    =e^{-\beta |k|}\langle \gamma_p \gamma_k^\dagger  \rangle_\beta 
    =e^{-\beta |k|}  (-\langle \gamma_k^\dagger \gamma_p\rangle_\beta +\delta_{kp})\ .
\end{equation}
At the first equality, we used Eq.~(\ref{gammaformula2}). Therefore, we have
\begin{equation}
    \langle \gamma_k^\dagger \gamma_p\rangle_\beta = \frac{\delta_{kp}}{e^{\beta |k|}+1}\ .
    \label{gg1}
\end{equation}
This is nothing but the Fermi distribution function. In a similar way, we also have
\begin{equation}
    \langle \gamma_k \gamma_p^\dagger \rangle_\beta = \frac{\delta_{kp}}{e^{-\beta |k|}+1}\ .
     \label{gg2}
\end{equation}

Now, we address the thermal Wightman function~(\ref{thermalWight}). Then, terms involving four $\gamma$'s 
appear. These terms can be computed as, for example, 
\begin{equation}
\begin{split}
    \langle\gamma_{-k'}^\dagger \gamma_k \gamma_p^\dagger \gamma_{-p'} \rangle_\beta
    &=e^{-\beta k'}\langle\gamma_k \gamma_p^\dagger \gamma_{-p'} \gamma_{-k'}^\dagger  \rangle_\beta \\
    &=e^{-\beta k'}\langle-\gamma_{-k'}^\dagger \gamma_k \gamma_p^\dagger \gamma_{-p'} +\delta_{k'p'} \gamma_k \gamma_p^\dagger\rangle_\beta  \ .
\end{split}
\end{equation}
At the first equality, we have used Eq.~(\ref{gammaformula2}).  At the second equality, we have moved $\gamma_{-k'}^\dagger$ to the front of the sequence of operators using the anti-commutation relation~(\ref{gammacomu}). (Note that all wavenumbers $k,k',p,p'$ are positive.)
Thus, together with Eq.~(\ref{gg2}), we obtain 
\begin{equation}
    \langle\gamma_{-k'}^\dagger \gamma_k \gamma_p^\dagger \gamma_{-p'} \rangle_\beta=\frac{\delta_{kp} \delta_{k'p'} }{(e^{-\beta k}+1)(e^{\beta k'}+1)}\ .
\end{equation}
Similarly, we also have 
\begin{align}
     &\langle\gamma_{k} \gamma_{-k'} \gamma_{-p'}^\dagger \gamma_p^\dagger \rangle_\beta
    =\frac{\delta_{kp} \delta_{k'p'}}{(e^{-\beta k}+1)(e^{-\beta k'}+1)}\ ,\\
    &\langle \gamma_{-k'}^\dagger \gamma_k^\dagger  \gamma_{p} \gamma_{-p'} \rangle_\beta
    =\frac{\delta_{kp} \delta_{k'p'}}{(e^{\beta k}+1)(e^{\beta k'}+1)}\ ,\\
    &\langle \gamma_k^\dagger \gamma_{-k'}  \gamma_{-p'}^\dagger \gamma_p \rangle_\beta
    =\frac{\delta_{kp} \delta_{k'p'}}{(e^{\beta k}+1)(e^{-\beta k'}+1)}\ .
\end{align}
As a result, the thermal Wightman function is given by
\begin{equation}
\begin{split}
&G^+_\beta(t,x,t',x')\\
&=\frac{1}{\ell^2}\sum_{k,k'\in K^+}\bigg[
\frac{e^{-ik(u-u')+ik'(v-v')}}{(e^{-\beta k}+1)(e^{\beta k'}+1)}
+\frac{e^{-ik(u-u')-ik'(v-v')}}{(e^{-\beta k}+1)(e^{-\beta k'}+1)}\\
&\hspace{3cm}+\frac{e^{ik(u-u')+ik'(v-v')}}{(e^{\beta k}+1)(e^{\beta k'}+1)}
+\frac{e^{ik(u-u')-ik'(v-v')}}{(e^{\beta k}+1)(e^{-\beta k'}+1)}\bigg]\\
&=g_\beta(u-u') g_\beta(v-v')\ ,
\end{split}
\end{equation}
where 
\begin{equation}
\begin{split}
    &g_\beta(u)\equiv \frac{1}{\ell}\sum_{k\in K^+}\left(\frac{e^{iku}}{e^{\beta k}+1}+\frac{e^{-ik u}}{e^{-\beta k}+1}\right)
    =\sum_{k\in K^+}\frac{e^{iku-\beta k}+e^{-iku}}{1+e^{-\beta k}}\\
    &=\frac{1}{\ell}\sum_{n=0}^\infty (-1)^n \sum_{k\in K^+}(e^{-i k(-u-i(n+1)\beta)}+e^{-ik(u-in\beta)})\\
    &=\sum_{n=0}^\infty (-1)^n (g(-u-i(n+1)\beta)+g(u-in\beta))\\
    &=\sum_{n=-\infty}^\infty (-1)^n g(u-in\beta)\ .
    \end{split}
\end{equation}
Here we use the relation $g(-u+in\beta)=-g(u-in\beta)$ for $n\neq0$.
In the limit of $\ell\to \infty$, 
\begin{equation}
\begin{split}
    g_\beta(u)|_{\ell\to \infty} &= \frac{1}{2\pi i} \sum_{n=-\infty}^\infty \frac{(-1)^n}{u -in\beta -i\epsilon}
    =\frac{1}{2\pi i} \left(\sum_{n=1}^\infty \frac{2(-1)^n u}{u^2+n^2\beta^2} + \frac{1}{u-i\epsilon}\right)\\
    &=\frac{1}{2i \beta \sinh\left[\frac{\pi}{\beta} (u-i\epsilon)\right]} .
\end{split}
\end{equation}
At the last equality, we have used
\begin{equation}
    \sum_{n=1}^\infty \frac{(-1)^n}{x^2+n^2}=-\frac{1}{2x^2}+\frac{\pi}{2x\sinh \pi x} \ .
\end{equation}
Therefore, the thermal Wightman function is given by
\begin{equation}
    G^+_\beta(t,x,t',x')|_{\ell\to \infty} = -\frac{1}{4\beta^2 \sinh\left[\frac{\pi}{\beta} (u-u'-i\epsilon)\right] \sinh\left[\frac{\pi}{\beta} (v-v'-i\epsilon)\right]}\ .
\end{equation}

\section{Diagnalization of the transverse-field Ising model}
\label{TFI}

In this appendix, we diagonalize the transverse-field Ising model. (See also Refs.~\cite{Mbeng,Molignini} for nice reviews.)  Using the results, we calculate the Wightman functions in the transverse-field Ising model. We apply the Fourier transformation of the operator $c_j$ and $c_j^\dagger$ as 
\begin{equation}
\begin{split}
 &c_j = \frac{1}{\sqrt{L}}\sum_{\kappa \in \mathcal{K}} e^{i \kappa j} c_\kappa\ ,\quad 
 c_j^\dagger = \frac{1}{\sqrt{L}}\sum_{\kappa \in \mathcal{K}} e^{i \kappa j} c^\dagger_{-\kappa}\ ,\\
 &\mathcal{K}=\left\{\frac{2\pi}{L} \left(n-\frac{1}{2}\right) \, \middle| \, n=-\frac{L}{2}+1,\dots,\frac{L}{2}\right\}\ .
 \end{split}
\end{equation}
Substituting the above expressions into Eq.~(\ref{Hforc}), we obtain the Hamiltonian in the momentum space as 
\begin{equation}
\begin{split}
    \hat{H}_\textrm{Spin}&=-\frac{1}{2\varepsilon}\sum_{\kappa\in \mathcal{K}} [e^{-i\kappa} c_{-\kappa} c_{\kappa} + e^{i\kappa} c_{\kappa}^\dagger c_{-\kappa}^\dagger  + 2\cos \kappa c_{\kappa}^\dagger c_{\kappa} + (1-2c_{\kappa}^\dagger c_{\kappa})]\\
    &=\frac{1}{\varepsilon}\sum_{\kappa\in \mathcal{K}^+} [
    (1-\cos\kappa) (c_{\kappa}^\dagger c_{\kappa} - c_{-\kappa}c_{-\kappa}^\dagger )
    -i\sin\kappa \, (c_{\kappa}^\dagger c_{-\kappa}^\dagger - c_{-\kappa}   c_{\kappa}) 
    ]\\
    &=\sum_{\kappa\in \mathcal{K}^+} \pmat{c_\kappa^\dagger & c_{-\kappa}}
    M \pmat{c_\kappa \\ c_{-\kappa}^\dagger}\ ,
\end{split}
\end{equation}
where $\mathcal{K}^+=\{\kappa\in \mathcal{K} \mid \kappa>0\}$ and the matrix $M$ is defined as 
\begin{equation}
    M = \pmat{z_\kappa & -iy_\kappa \\ iy_\kappa & -z_\kappa }\ ,
    \quad z_\kappa=\frac{1-\cos \kappa}{\varepsilon}\ ,\quad y_\kappa=\frac{\sin \kappa}{\varepsilon}\ .
\end{equation}
This matrix is diagonalized as
\begin{equation}
    U^\dagger M U = \pmat{\epsilon_\kappa & 0 \\ 0 & -\epsilon_\kappa }\ ,\quad 
    U=\pmat{u_\kappa & -v_\kappa^\ast\\ v_\kappa & u_\kappa^\ast}\ ,
\end{equation}
where 
\begin{equation}
\begin{split}
    \epsilon_\kappa=\sqrt{z_\kappa^2+y_\kappa^2}=\frac{2}{\varepsilon}\left|\sin\frac{\kappa}{2}\right|\ ,\quad  \pmat{u_\kappa \\ v_\kappa} = \frac{1}{\sqrt{2\epsilon_\kappa(\epsilon_\kappa+z_\kappa)}}\pmat{\epsilon_\kappa+z_\kappa \\ iy_\kappa}\ .
\end{split}
\end{equation}
Thus, defining operator $\gamma_{\kappa}$ as
\begin{equation}
\pmat{\gamma_\kappa \\ \gamma_{-\kappa}^\dagger} = U^\dagger \pmat{c_\kappa \\ c_{-\kappa}^\dagger}\ ,
\end{equation}
we have the diagonalized Hamiltonian as
\begin{equation}
    \hat{H}_\textrm{Spin}=\sum_{\kappa\in \mathcal{K}^+} \epsilon_\kappa(\gamma_\kappa^\dagger \gamma_{\kappa}-\gamma_{-\kappa} \gamma_{-\kappa}^\dagger)
    =\sum_{\kappa\in \mathcal{K}} \epsilon_\kappa \gamma_\kappa^\dagger \gamma_{\kappa} + E_0\ ,
\end{equation}
where $E_0=-\sum_{\kappa\in \mathcal{K}} \epsilon_\kappa/2$.
The ground state is defined by
\begin{equation}
    \gamma_\kappa |\Omega\rangle =0\quad (\forall \kappa \in \mathcal{K})\ .
    \label{Omegadef}
\end{equation}
In the Heisenberg picture, $\gamma_\kappa(\eta)=\gamma_\kappa(0)e^{-i\epsilon_\kappa \eta}$. Therefore, the solution of Heisenberg equation written in the position space is 
\begin{equation}
\begin{split}
    c_j(\eta)&=\frac{1}{\sqrt{L}}\sum_{\kappa \in \mathcal{K}} e^{i\kappa j} c_\kappa(\eta) = \frac{1}{\sqrt{L}}\sum_{\kappa \in \mathcal{K}} e^{i\kappa j}(u_\kappa \gamma_\kappa e^{-i\epsilon_\kappa \eta} + v_\kappa \gamma_{-\kappa}^\dagger e^{i\epsilon_\kappa \eta})\ ,\\
     c_j^\dagger(\eta)&=\frac{1}{\sqrt{L}}\sum_{\kappa \in \mathcal{K}} e^{i\kappa j} c_\kappa(\eta) = \frac{1}{\sqrt{L}}\sum_{\kappa \in \mathcal{K}} e^{i\kappa j}(v_\kappa \gamma_\kappa e^{-i\epsilon_\kappa \eta} + u_\kappa \gamma_{-\kappa}^\dagger e^{i\epsilon_\kappa \eta})\ ,
    \end{split}
\end{equation}
where $\gamma_\kappa=\gamma_\kappa(0)$. 

We define the number operator for the fermion operator at the $j$-th site as
\begin{equation}
    n_j(\eta)=\nord{c_j(\eta)^\dagger c_j(\eta)} = \nord{\frac{1}{2}(1-\sigma^z_j(\eta))}=-\frac{1}{2}(\sigma^z_j(\eta)-\langle\sigma^z_j(\eta)\rangle)\ .
\end{equation}
We will compute the Wightman function for this operator.
Using the definition of the ground state~(\ref{Omegadef}), one can compute
\begin{equation}
    n_j(\eta) |\Omega\rangle =
    \frac{1}{L}\sum_{\kappa \kappa '\in \mathcal{K}}u_{\kappa '}v_{\kappa }e^{i(\epsilon_{\kappa }+\epsilon_{\kappa '})\eta+i(\kappa +\kappa ')j}|-\kappa ',-\kappa \rangle\ ,
\end{equation}
where 
\begin{equation}
    |-\kappa ,-\kappa '\rangle =\gamma_{-\kappa '}^\dagger\gamma_{-\kappa}^\dagger |\Omega\rangle \ .
\end{equation}
We also have
\begin{equation}
     \langle \Omega | n_j(\eta)
    =
    -\frac{1}{L}\sum_{\kappa \kappa '\in \mathcal{K}}u_{\kappa '}v_{\kappa }e^{-i(\epsilon_{\kappa }+\epsilon_{\kappa '})\eta-i(\kappa +\kappa ')j}\langle -\kappa ,-\kappa '|\ ,
\end{equation}
where 
\begin{equation}
    \langle -\kappa ,-\kappa '|=\langle \Omega | \gamma_{-\kappa} \gamma_{-\kappa '}\ .
\end{equation}
We can check
\begin{equation}
    \langle -p,-p'|-\kappa ',-\kappa \rangle = \delta_{p\kappa }\delta_{p'\kappa '}-\delta_{p\kappa '}\delta_{p'\kappa }\ .
\end{equation}
Therefore, the Wightman function for $n_j(\eta)$ is given by
\begin{equation}
    \langle n_{j}(\eta)n_{j'}(\eta')\rangle
    =-\frac{1}{L^2}\sum_{\kappa \kappa '\in \mathcal{K}}u_{\kappa '}v_{\kappa }(u_{\kappa '}v_{\kappa }-u_{\kappa }v_{\kappa '})
    e^{-i(\epsilon_{\kappa }+\epsilon_{\kappa '})\Delta \eta-i(\kappa +\kappa ')\Delta j}\ ,
\end{equation}
where $\Delta \eta = \eta-\eta'$ and $\Delta j = j-j'$.

\bibliography{refs}

\providecommand{\href}[2]{#2}\begingroup\raggedright\begin{thebibliography}{10}

\bibitem{Gibbons:1977mu}
G.~W. Gibbons and S.~W. Hawking, \emph{{Cosmological Event Horizons, Thermodynamics, and Particle Creation}}, \href{https://doi.org/10.1103/PhysRevD.15.2738}{\emph{Phys. Rev. D} {\bfseries 15} (1977) 2738}.

\bibitem{Hotta:2022aiv}
M.~Hotta, Y.~Nambu, Y.~Sugiyama, K.~Yamamoto and G.~Yusa, \emph{{Expanding edges of quantum Hall systems in a cosmology language: Hawking radiation from de Sitter horizon in edge modes}}, \href{https://doi.org/10.1103/PhysRevD.105.105009}{\emph{Phys. Rev. D} {\bfseries 105} (2022) 105009} [\href{https://arxiv.org/abs/2202.03731}{{\ttfamily 2202.03731}}].

\bibitem{Nambu:2023tpg}
Y.~Nambu and M.~Hotta, \emph{{Analog de Sitter universe in quantum Hall systems with an expanding edge}}, \href{https://doi.org/10.1103/PhysRevD.107.085002}{\emph{Phys. Rev. D} {\bfseries 107} (2023) 085002} [\href{https://arxiv.org/abs/2301.09270}{{\ttfamily 2301.09270}}].

\bibitem{Unruh1}
W.~G. Unruh, \emph{Notes on black-hole evaporation}, \href{https://doi.org/10.1103/PhysRevD.14.870}{\emph{Phys. Rev. D} {\bfseries 14} (1976) 870}.

\bibitem{DeWitt1}
B.~S. DeWitt, \emph{General relativity: An einstein centenary survey, edited by s. w. hawking and w. israel}, {\emph{Cambridge University Press, Cambridge} 680}.

\bibitem{BD}
N.~Birrell and P.~Davies, \emph{{Quantum Fields in Curved Space.}}, {\emph{Cambridge University Press} (1982) }.

\bibitem{Bousso:2001mw}
R.~Bousso, A.~Maloney and A.~Strominger, \emph{{Conformal vacua and entropy in de Sitter space}}, \href{https://doi.org/10.1103/PhysRevD.65.104039}{\emph{Phys. Rev. D} {\bfseries 65} (2002) 104039} [\href{https://arxiv.org/abs/hep-th/0112218}{{\ttfamily hep-th/0112218}}].

\bibitem{Kinoshita:2024ahu}
S.~Kinoshita, K.~Murata, D.~Yamamoto and R.~Yoshii, \emph{{Spin systems as quantum simulators of quantum field theories in curved spacetimes}},  \href{https://arxiv.org/abs/2410.07587}{{\ttfamily 2410.07587}}.

\bibitem{Hummer:2015xaa}
D.~H\"ummer, E.~Martin-Martinez and A.~Kempf, \emph{{Renormalized Unruh-DeWitt Particle Detector Models for Boson and Fermion Fields}}, \href{https://doi.org/10.1103/PhysRevD.93.024019}{\emph{Phys. Rev. D} {\bfseries 93} (2016) 024019} [\href{https://arxiv.org/abs/1506.02046}{{\ttfamily 1506.02046}}].

\bibitem{Kubo}
R.~Kubo, \emph{Statistical-mechanical theory of irreversible processes. i. general theory and simple applications to magnetic and conduction problems}, \href{https://doi.org/10.1143/JPSJ.12.570}{\emph{Journal of the Physical Society of Japan} {\bfseries 12} (1957) 570} [\href{https://arxiv.org/abs/https://doi.org/10.1143/JPSJ.12.570}{{\ttfamily https://doi.org/10.1143/JPSJ.12.570}}].

\bibitem{MS}
P.~C. Martin and J.~Schwinger, \emph{Theory of many-particle systems. i}, \href{https://doi.org/10.1103/PhysRev.115.1342}{\emph{Phys. Rev.} {\bfseries 115} (1959) 1342}.

\bibitem{Polyakov:1981re}
A.~M. Polyakov, \emph{{Quantum Geometry of Fermionic Strings}}, \href{https://doi.org/10.1016/0370-2693(81)90744-9}{\emph{Phys. Lett. B} {\bfseries 103} (1981) 211}.

\bibitem{deLacroix:2023uem}
C.~de~Lacroix, H.~Erbin and V.~Lahoche, \emph{{Gravitational action for a massive Majorana fermion in 2d quantum gravity}}, \href{https://doi.org/10.1007/JHEP01(2024)068}{\emph{JHEP} {\bfseries 01} (2024) 068} [\href{https://arxiv.org/abs/2308.08342}{{\ttfamily 2308.08342}}].

\bibitem{Erbin}
H.~Erbin, \emph{Gravitational action for a massive majorana fermion in 2d quantum gravity ^^e2^^80^^93 notes},  2021.

\bibitem{Bunch:1978yq}
T.~S. Bunch and P.~C.~W. Davies, \emph{{Quantum Field Theory in de Sitter Space: Renormalization by Point Splitting}}, \href{https://doi.org/10.1098/rspa.1978.0060}{\emph{Proc. Roy. Soc. Lond. A} {\bfseries 360} (1978) 117}.

\bibitem{Brandenberger:1999sw}
R.~H. Brandenberger, \emph{{Inflationary cosmology: Progress and problems}},  in \emph{{IPM School on Cosmology 1999: Large Scale Structure Formation}}, 1, 1999, \href{https://arxiv.org/abs/hep-ph/9910410}{{\ttfamily hep-ph/9910410}}.

\bibitem{Daley2022-ok}
A.~J. Daley, I.~Bloch, C.~Kokail, S.~Flannigan, N.~Pearson, M.~Troyer et~al., \emph{Practical quantum advantage in quantum simulation}, {\emph{Nature} {\bfseries 607} (2022) 667}.

\bibitem{Devoret2013-bg}
M.~H. Devoret and R.~J. Schoelkopf, \emph{Superconducting circuits for quantum information: an outlook}, {\emph{Science} {\bfseries 339} (2013) 1169}.

\bibitem{Wendin2017-io}
G.~Wendin, \emph{Quantum information processing with superconducting circuits: a review}, {\emph{Rep. Prog. Phys.} {\bfseries 80} (2017) 106001}.

\bibitem{Houck2012-hc}
A.~A. Houck, H.~E. T{\"u}reci and J.~Koch, \emph{On-chip quantum simulation with superconducting circuits}, {\emph{Nat. Phys.} {\bfseries 8} (2012) 292}.

\bibitem{Kim2023-nz}
Y.~Kim, A.~Eddins, S.~Anand, K.~X. Wei, E.~van~den Berg, S.~Rosenblatt et~al., \emph{Evidence for the utility of quantum computing before fault tolerance}, {\emph{Nature} {\bfseries 618} (2023) 500}.

\bibitem{Barreiro2011-cp}
J.~T. Barreiro, M.~M{\"u}ller, P.~Schindler, D.~Nigg, T.~Monz, M.~Chwalla et~al., \emph{An open-system quantum simulator with trapped ions}, {\emph{Nature} {\bfseries 470} (2011) 486}.

\bibitem{Blatt2012-fa}
R.~Blatt and C.~F. Roos, \emph{Quantum simulations with trapped ions}, {\emph{Nat. Phys.} {\bfseries 8} (2012) 277}.

\bibitem{Monroe2021-tn}
C.~Monroe, W.~C. Campbell, L.-M. Duan, Z.-X. Gong, A.~V. Gorshkov, P.~W. Hess et~al., \emph{Programmable quantum simulations of spin systems with trapped ions}, {\emph{Rev. Mod. Phys.} {\bfseries 93} (2021) }.

\bibitem{Weimer2010-fu}
H.~Weimer, M.~M{\"u}ller, I.~Lesanovsky, P.~Zoller and H.~P. B{\"u}chler, \emph{A rydberg quantum simulator}, {\emph{Nat. Phys.} {\bfseries 6} (2010) 382}.

\bibitem{Henriet2020-yk}
L.~Henriet, L.~Beguin, A.~Signoles, T.~Lahaye, A.~Browaeys, G.-O. Reymond et~al., \emph{Quantum computing with neutral atoms}, {\emph{Quantum} {\bfseries 4} (2020) 327}.

\bibitem{Mbeng}
G.~E.~S. Glen Bigan~Mbeng, Angelo~Russomanno, \emph{{The quantum Ising chain for beginners}},  \href{https://arxiv.org/abs/2009.09208}{{\ttfamily 2009.09208}}.

\bibitem{Molignini}
P.~Molignini, \emph{{Analyzing the two dimensional Ising model with conformal fieldtheory}}, {\emph{Report of Proseminar on Conformal Field Theory and String Theory ^^e2^^80^^93 FS 2013} (2013) }.

\end{thebibliography}\endgroup

\end{document}